\documentclass[notitlepage,aps,letter,reprint,footinbib]{revtex4-1}

\usepackage{graphicx}
\usepackage{amsmath}
\usepackage{amssymb}
\usepackage{hyperref}
\usepackage{tikz}
\usepackage[english]{babel}

\newcommand{\be}{\begin{eqnarray*}}
\newcommand{\ee}{\end{eqnarray*}}
\newcommand{\beq}{\begin{eqnarray}}
\newcommand{\eeq}{\end{eqnarray}}
\newcommand{\bequ}{\begin{equation}}
\newcommand{\eequ}{\end{equation}}

\newcommand{\br}{{\mathbf{r}}}

\newcommand{\dd}{\mathrm{d}}

\newcommand{\h}{\hat{H}}

\newcommand{\ket}[1]{\left|{#1}\right\rangle}
\newcommand{\bra}[1]{\left\langle{#1}\right|}

\newcommand{\zbb}{\mathbb{Z}}

\newcommand{\id}{\mathbb{I}}

\begin{document}
\title{Floquet topological phases with symmetry in all dimensions}
\author{Rahul Roy}
\author{Fenner Harper}
\affiliation{Department of Physics and Astronomy, University of California, Los Angeles, California 90095, USA}
\date{\today}
\begin{abstract}
Dynamical systems may host a number of remarkable symmetry-protected phases that are qualitatively different from their static analogs. In this work, we consider the phase space of symmetry-respecting unitary evolutions in detail and identify several distinct classes of evolution that host novel dynamical order. Using ideas from group cohomology, we construct a set of interacting Floquet drives that generate dynamical symmetry-protected topological order for each nontrivial cohomology class in every dimension, illustrating our construction with explicit two-dimensional examples. We also identify a set of symmetry-protected Floquet drives that lie outside of the group cohomology construction, and a further class of symmetry-respecting topological drives which host chiral edge modes. We use these special drives to define a notion of phase (stable to a class of local perturbations in the bulk) and the concepts of relative and absolute topological order, which can be applied to many different classes of unitary evolutions. These include fully many-body localized unitary evolutions and time crystals.  

\end{abstract}
\maketitle

\section{Introduction}

Driven systems have recently been shown to support a remarkable set of dynamical phases and phenomena for which there is no static analog. In this rapidly evolving area, there has been a particular focus on novel topological phases, whose boundaries may exhibit exotic dynamical edge modes. Notably, a unified classification of Floquet topological insulators has recently been obtained \cite{Roy:2016via}, and much progress has been made towards understanding Floquet symmetry-protected topological phases (FSPTs) in one dimension and higher \cite{Khemani:2016gd,vonKeyserlingk:2016bq,vonKeyserlingk:2016ea,Else:2016ja,Potter:2016tba,Roy:2016ka,Potter:2016vq}. More recently, systems have been proposed that break discrete translational symmetry in the time domain (dubbed `time crystals' or `$\pi$-spin glasses') \cite{Khemani:2016gd,vonKeyserlingk:2016ea,Else:2016wx,vonKeyserlingk:2016ev,Else:2016gf,Yao:2016wpa}, and dynamical two-dimensional models have been obtained which possess a form of topological order that does not depend on symmetry, and which is manifested as anomalous chiral edge modes \cite{Po:2016uwb,Harper:2016us}. Many of these exciting new phases are well suited to experimental realization: topological Floquet states have been observed in photonic systems \cite{Kitagawa:2012gl,Rechtsman:2013fe,Maczewsky:2016ua} and using cold atoms \cite{Jotzu:2015kz,JimenezGarcia:2015kd}, and discrete time crystals have also recently been realised \cite{Zhang:2016uw,Choi:2016wn}.

In this paper, we focus on novel bosonic Floquet systems with symmetry and strong interactions, which we find may be entirely characterized by their edge behavior. This is a form of holography peculiar to driven systems that is analogous to the bulk-edge connection of static symmetry-protected topological phases (SPTs) (see Refs.~\cite{Chen:2013foa,Senthil:2015ci,Zeng:2015vc} and references therein). For example, in one dimension, FSPT phases are characterized by irreducible representations of their protecting symmetry group, which are manifested at the ends of an open chain \cite{vonKeyserlingk:2016bq,Else:2016ja,Roy:2016ka,Potter:2016tba}. We find in this paper that, in all dimensions, drives may be constructed which map product states at a boundary onto nontrivial SPT states. This is an explicit realization in all dimensions of the idea of pumping proposed in one and two dimensions in Ref.~\cite{vonKeyserlingk:2016bq,Potter:2016tba,Else:2016ja}. Very recently, after this work was largely complete, the pumping approach was also explored in a set of Abelian models in two dimensions in Ref.~\cite{Potter:2016vq}. In past work~\cite{Else:2016ja,Potter:2016tba}, FSPT phases were conjectured to have a group cohomology classification. Our work proves that cohomology classes define distinct FSPT phases, which is indeed consistent with this conjecture, although there are some important differences in our approach that are discussed later. 

The nontrivial drives that we construct capture an inherently dynamical kind of SPT order, which may be characterized by an `effective edge unitary', the action of the evolution at the boundary of a $d$-dimensional open system. Remarkably, these $(d-1)$-dimensional effective edge unitaries cannot be generated in a symmetry-preserving $(d-1)$-dimensional system, and so are topologically robust to local perturbations. We also discuss a qualitatively different set of symmetry-protected drives that are unique to two dimensions, and which are based on the (non-symmetry protected) anomalous chiral drives of Refs.~\cite{Po:2016uwb,Harper:2016us}. These are distinct from the FSPT phases constructed based on group cohomology, and do not seem to have been anticipated in the literature.

The notion of a phase in Floquet systems is a subtle and delicate one: While a particular model drive may have exotic topological properties, it is far harder to demonstrate that this model is representative of a wider class of models. To extend these special topological drives to Floquet phases requires a careful consideration of phase robustness, demonstrating stability to weak perturbations and heating effects that would inevitably arise in a generic interacting, driven system. For this reason, we spend the first part of this article discussing the phase structure of dynamical systems in general. 

To aid the discussion of phase robustness, we will define families of unitary evolutions whose endpoints satisfy certain properties. For example, a family of unitary evolutions could share the same, fixed endpoint. Within such a family, one can define the relative order between any two members and, in turn, define an analog of phase based on this relative order. A particularly useful notion of phase may be developed for endpoint unitaries which possess a complete set of local integrals of motion, such as in many-body localized (MBL) systems \cite{Nandkishore:2015kt}. In the static case, in the presence of MBL, ground state order (such as SPT order, topological order or spontaneous symmetry breaking) can become a property of the entire spectrum, and is then protected in the absence of a delocalization transition \cite{Huse:2013bw,Bauer:2013jw,Chandran:2014dk,Slagle:2015uo,Potter:2015vn,Bahri:2015ib}. In addition, MBL protects against heating effects \cite{DAlessio:2013fv,Lazarides:2014ie,Abanin:2014te,Ponte:2015hm,Ponte:2015dc,Lazarides:2015jd,Abanin:2015bc,Khemani:2016gd,Zhang:2016vt,Zhang:2016tb} and may assist in the detection of edge modes \cite{Chandran:2014dk,Bahri:2015ib}, and has recently been realized in a variety of experimental settings \cite{Schreiber:2015jt,Smith:2016cd,Choi:2016ic,Bordia:2016tw}. While the existence of MBL in dimensions greater than one remains a matter of conjecture (and indeed, some doubt has recently been cast on this conjecture \cite{DeRoeck:2016us,Agarwal:2016vk}), such phases are expected to be robust on timescales that are exponentially large in the disorder strength. 

In this paper, as in some of our previous work \cite{Roy:2016ka,Roy:2016via,Harper:2016us}, we disentangle the discussion of MBL from the topological aspects of dynamical systems, which can be studied independently. We discuss the phase structure of dynamical systems in terms of the space of unitary evolution operators, and use ideas from homotopy to classify phases through the connectedness of this space. The advantage of our approach is that we can discuss the topological aspects of various types of nonthermal unitaries in a unified manner and can separate the more rigorously understood topological aspects of the study from the less well-understood localization-related aspects. Thus, the discussion may be easily extended to time crystals, for example, as well as systems with only partial MBL. Our study makes use of drives that we call loops, which have the specific property that the Floquet Hamiltonian is effectively zero in a closed system, and is thus somewhat different from other approaches which have focused on bulk eigenstate order. 

The structure of the paper is as follows. In Section~\ref{sec:phase_structure_u}, we discuss the phase space of unitary evolutions in detail and explain the homotopic ideas that motivate the unitary loop models we use throughout the document. These ideas are explored further in Sec.~\ref{sec:nontrivial_unitary_loops}, where we discuss some properties of loops and the different types of loop order that may arise. In Sec.~\ref{sec:TSDs} we construct a set of symmetric topological drives, based on the construction of Refs.~\cite{Po:2016uwb,Harper:2016us}, which are nontrivial even in the absence of symmetry restrictions. These motivate a set of FSPT drives introduced in Sec.~\ref{sec:FSPT_not_cohomology} that are outside of the group cohomology paradigm. The bulk of the paper then uses a group cohomology approach to construct a general set of FSPT drives applicable to any unitary symmetry group $G$ in any dimension. We introduce the construction in two dimensions (with examples) in Sec.~\ref{sec:cohomology_2d}, before generalizing to all dimensions in Sec.~\ref{sec:cohomology_anyd}. In Sec.~\ref{sec:W_standardpaths} we discuss the utility of our nontrivial drives in describing order in generic Floquet systems. Finally, we summarize our results and make some concluding remarks in Sec.~\ref{sec:conclusion}. Some of the less essential proofs and discussions are omitted from the main text and may be found in the Appendices. We also provide a detailed comparison between our approach to the study of FSPT phases and other approaches based on eigenstate order in Appendix~\ref{sec:eigenstate_order}.

\section{Phase Structure of Unitary Evolutions\label{sec:phase_structure_u}}
Throughout this paper, we are interested in classifying the types of stable order that can exist in the space of unitary evolutions. With this aim in mind, we begin by discussing some general concepts related to the nature of this space and the subtle notion of a phase within it. While the discussion will initially be general (and fairly abstract), we will later place these ideas in a more concrete setting.
\subsection{Static Systems}

It is useful to first recall the notion of a  gapped quantum phase in a static system. A particular gapped Hamiltonian may be thought of as belonging to a continuous space of gapped Hamiltonians, $S$, which we may restrict by insisting that the Hamiltonians in the space obey a particular set of symmetries and possibly  have some other shared features. (For instance, we may require that they be noninteracting). We regard two gapped quantum Hamiltonians as belonging to distinct phases only if there is no way to adiabatically transform one Hamiltonian to the other within this space (i.e. without closing the gap or breaking any symmetry). A classification of static quantum phases is therefore a statement about the basic topology (connectedness) of the space $S$. 

In the presence of localization, specifically MBL for interacting systems \cite{Nandkishore:2015kt}, the ordering properties of a given gapped quantum phase may be shared by all or a large fraction of the eigenstates of the Hamiltonian, rather than just the ground state \cite{Huse:2013bw,Bauer:2013jw,Chandran:2014dk,Slagle:2015uo,Potter:2015vn,Bahri:2015ib}. The order of these eigenstates can then be protected by the conjectured stability of MBL to local perturbations, and cannot change in the absence of a delocalization transition.  This permits a definition of a distinct phase for MBL Hamiltonians with each type of order, without the need for a persistent gap. A classification of MBL phases therefore describes the topology of the space of MBL Hamiltonians, $S_{\rm MBL}$ (which may also have symmetry restrictions imposed upon it).
\subsection{General Concepts for Driven Systems}
Unitary evolutions are inherently more complicated than static systems, but in principle may also be classified by studying the topology of their (possibly restricted) phase space. In this paper we restrict the discussion to the space of unitary evolutions with no symmetries (which we write as $S_0$) and spaces of unitary evolutions which possess unitary symmetries belonging to a group $G$ (which we write as $S_G$).

Within these spaces, we may regard a unitary evolution as being generated by some time-dependent Hamiltonian, through
\beq
U(t)&=&\mathcal{T}\exp\left[-i\int_0^tH(t')\dd t'\right],
\eeq
where, throughout this article, we take $t\in[0,1]$. In this way, a generic unitary evolution traces out a path in the space $S_0$ or $S_G$, starting at the origin, $U(0)=\id$. If the space is $S_G$, then the generating Hamiltonian satisfies the property
\beq
V(g)H(t)V(g)^{-1}&=&H(t),
\eeq
for all $t$ and for all $g\in G$, where $V(g)$ is the global representation of the group element $g$. The unitary evolution satisfies the corresponding symmetry property
\beq
V(g)U(t)V(g)^{-1}=U(t).
\eeq
In the more general case, one could also consider antiunitary symmetries, which relate the unitary at time $t$ to the unitary at some other time $t'$.

The effects of a local perturbation may be absorbed into a continuous deformation of the unitary evolution. However, it is clear that no unitary evolution is stable to arbitrary local perturbations: Any unitary evolution $U_1(t)$ in the space can be continuously deformed to the trivial unitary evolution $U(t)=\id$  and therefore into any other $U_2(t)$. To obtain a meaningful classification, we must therefore impose further restrictions on the types of local perturbations, and thus on the space of unitary evolutions. In other words, just as in the static case, we must consider a restricted class of perturbations: a symmetry restriction alone is clearly not sufficient. 

A natural way of restricting the phase space of unitary evolutions is to demand that the endpoint take a certain form, i.e., we consider a family of unitary evolutions whose end points have some natural shared properties, and restrict ourselves to local perturbations which do not take us outside of this family. For example, we may require the unitary evolution to end at a particular point, $U(1)=P$, or we may require its end point to have the form $U(t)=\exp(-iH_{\rm MBL})$, where $H_{\rm MBL}$ is some MBL Hamiltonian. The set of local perturbations that keep the end point unitary fixed is rather limited, but still defines a family of unitary evolutions which lives in an infinite dimensional space.  With the end point unitaries restricted to those that have a complete set of local integrals of motion, we expect the inherent stability of MBL Hamiltonians to transfer to these driven systems, protecting them from local perturbations that do not drive the end point through a delocalization transition. We consider these specific cases further below, but for now we keep the discussion general and write the restricted space of endpoints as $W$.

If the space of endpoints $W$ is itself disconnected (for example, if it consists of MBL Hamiltonians with different SPT orders), then this automatically leads to a discontinuity in the space of possible unitary evolutions. This classification of endpoints is in some sense similar to the classification of static phases and leads to one kind of order in driven systems. However, there may in general exist inequivalent paths with the same endpoint in $W$, leading to an inherently dynamical kind of order with no static analog. This motivates the definition of a unitary loop: if we have two unitary evolutions $U_1$ and $U_2$, which end at the same point so that $U_1(1)=U_2(1)=P$, then the composition of unitary evolutions $U_2^{-1}\circ U_1$ ends at the identity. Explicitly, if $U_j$ is generated by Hamiltonian $H_j(t)$, the composition $U_2^{-1}\circ U_1$ is generated by the Hamiltonian
\beq
H(t)&=&\left\{\begin{array}{cc}
H_1(2t) & 0\leq t\leq1/2\\
-H_2(2t-1) & 1/2\leq t\leq1,
\end{array}\right.
\eeq
which runs one unitary evolution after the other, rescaled to the time interval $t\in[0,1]$. More generally, we define a unitary loop as any unitary evolution that, for a closed system, satisfies $U(0)=U(1)=\id$. Loops play an important role in the classification of the phase space of unitary evolutions, as we now show.
\subsection{Absolute Order}
\begin{figure}[t]
\begin{center}
\includegraphics[clip=true, trim = 0 0 120 0,scale=0.49]{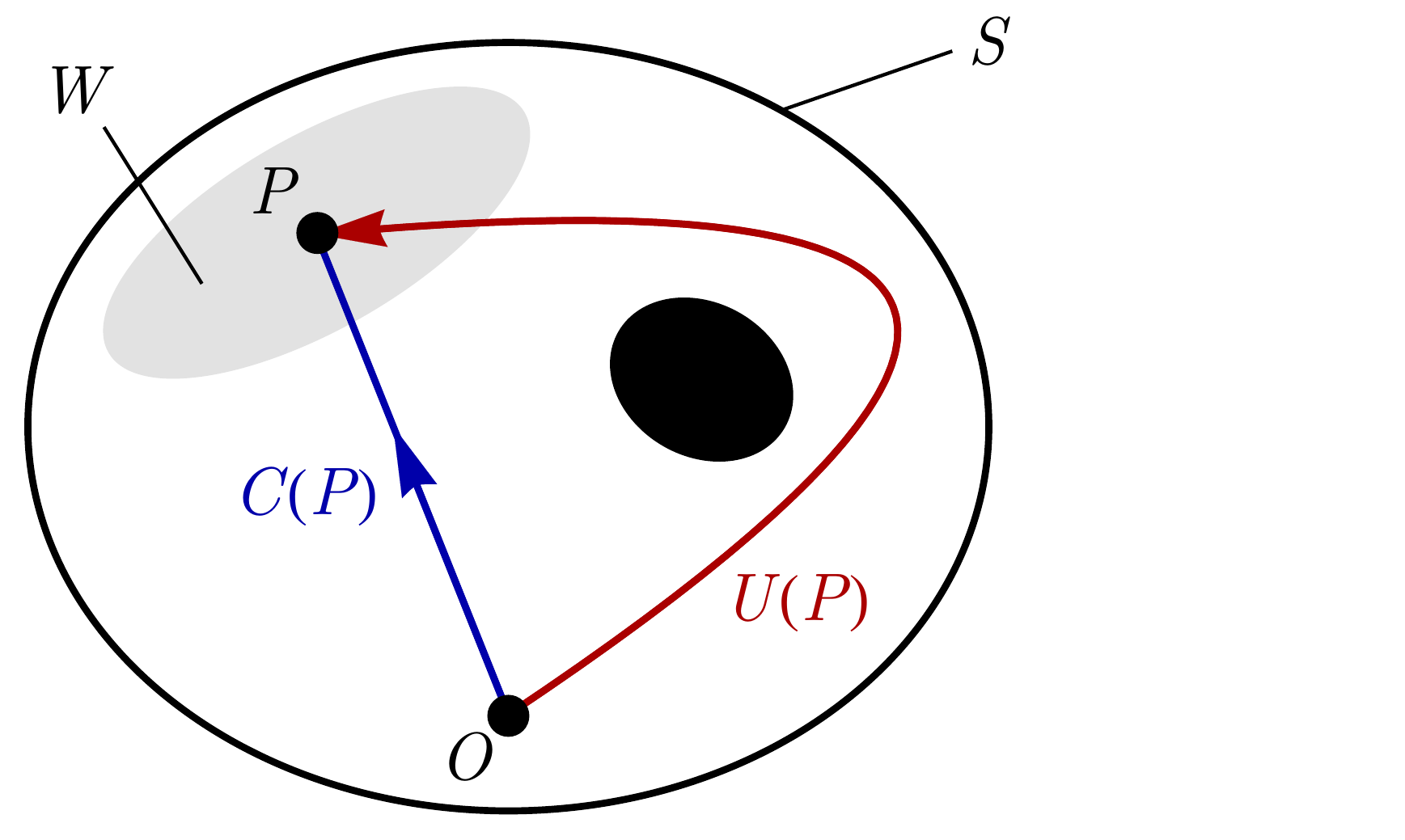}
\end{center}
\caption{Diagram of the space of unitary evolutions, $S$ (which may correspond to $S_0$ or some $S_G$). $W$ is a region of restricted endpoints to which unique paths may be defined and the black ellipse is a region that lies outside of $S$. $C(P)$ is the unique standard path to the point $P\in W$, and $U(P)$ is a different unitary evolution that ends at $P$. The composition $C(P)^{-1}\circ U(P)$ forms a noncontractible loop. \label{fig:homotopy1}}
\end{figure}

In general, we consider some (possibly symmetry-protected) space of unitary evolutions, $S\in\{S_0,S_G\}$, which contains some region of restricted endpoints, $W$, as shown in Fig.~\ref{fig:homotopy1}. Within $S$, we suppose there is a set of non-contractible loops, $\{L\}$, which are uniquely labeled by a collection of topological invariants that we collectively write as $n(L)$ (and sometimes refer to as the loop order). 

In order to facilitate definitions of order and phase for unitary evolutions, the region $W$ should be endowed with certain properties. First, suppose that for every point $P\in W$, there is a unique `standard path', $C(P)$, from the origin to the point $P$ (see below for examples of unitary evolutions for which this assumption is expected to hold). If this is true, then it follows that for any path $U(P)$ that ends at $P$, there is a unique unitary loop formed from $L(U)=C(P)^{-1}\circ U(P)$, where $\circ$ indicates the composition of paths (see Fig.~\ref{fig:homotopy1}). This allows a definition of absolute order for all evolutions $U(P)$ that end at $P$, labeled by the loop order, $n(L(U))$, associated with $L(U)$.

We may also relax the condition that there is a unique standard path to each point $P$, and assume instead that there is \emph{at least one} standard path to each point $P$, writing the collection of standard paths as $\{C_i(P)\}$. If every loops $C_j(P)^{-1}\circ C_i(P)$ is trivial, then choosing any standard path gives the same definition of absolute phase. If there is at least one nontrivial loop, then it may still be possible to define the notion of relative order.

\subsection{Relative Order\label{sec:rel_phase}}
Suppose we have two points in $W$, $P$ and $P'$, and are interested in comparing the drives associated with the paths $U(P)$ and $U'(P')$. We may ask if it is possible to deform $U(P)$ into $U'(P')$ through a set of local perturbations that keep the endpoint in $W$ throughout the deformation. Suppose that $S(P,P')$ is a path from $P$ to $P'$ contained entirely within region $W$. Then, if the loop order associated with $L(U,U')=U'(P')^{-1}\circ S(P,P')\circ U(P)$ has order $n$ for every path $S(P,P')$, then we may assert that $U(P)$ has loop order $n$ relative to $U'(P')$.

\begin{figure}[t]
\begin{center}
\includegraphics[clip=true, trim = 0 0 0 0,scale=0.5]{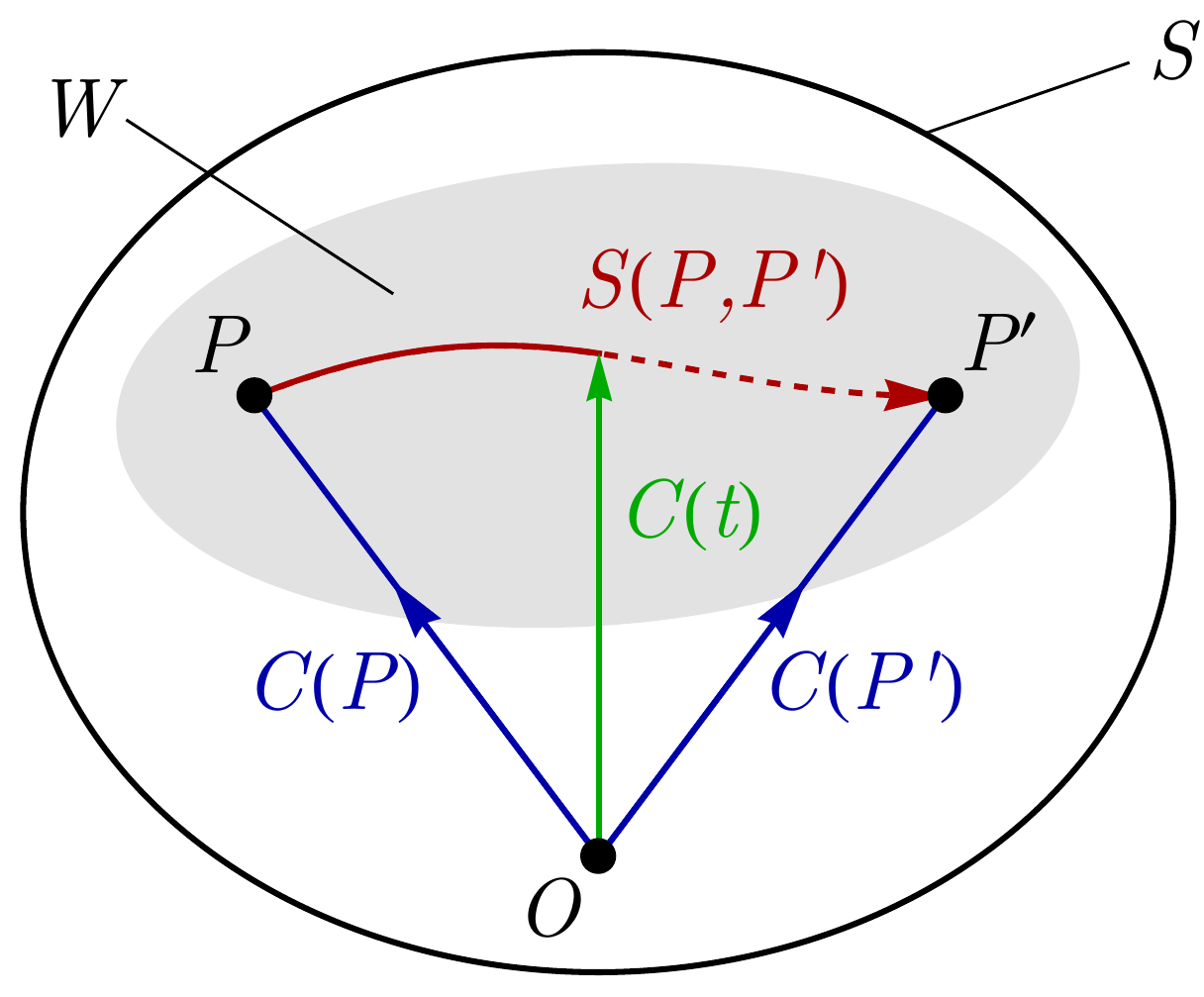}
\end{center}
\caption{Continuous path $S(P,P')$ between standard paths $C(P)$ and $C(P')$. If $C(t)$ is homotopic to $C(P)$, then $C(P')$ can be defined uniquely given $C(P)$. See main text for details. \label{fig:homotopy2}}
\end{figure}

We now reconsider the standard paths $\{C_i(P)\}$ introduced above and again consider the path $S(P,P')$ that connects $P$ to $P'$ and that is contained within $W$. Suppose this defines a continuous deformation of $C(P)$ along the path, i.e. that the path $C(t)=S(t)\circ C(P)$, where $S(t)$ is a parametrization of path $S(P,P')$, is homotopic to $C(P)$, as shown in Fig.~\ref{fig:homotopy2}. Then, we have a definition of $C(P')$ that depends on $C(P)$ and $S(P,P')$. If every such $S(P,P')$ gives an equivalent definition of $C(P')$, then there is a unique definition of $C(P')$ given $C(P)$. If this property holds for every $P'$, then choosing any $C_i(P)$ from $\{C_i(P)\}$ provides a unique definition of $C_i(P')$, and so we can use these to define a consistent notion of relative order. If this fails, then there are nontrivial loops in the parameter space.

\subsection{Spaces of Endpoint Unitaries}
From the discussions above, it is clear that a useful choice of $W$ (the restricted region of endpoint unitaries), is one in which there are no nontrivial loops. A simple example where this is true is when $W=P$ is a single point. In this case, the relative phase of any pair of unitary evolutions $U_1$, $U_2$ that end at $P$ can be obtained by calculating the loop order of the composition $U_2^{-1}\circ U_1$. There are other nontrivial choices of $W$, but a complete discussion of these builds on our construction of noncontractible loops. We therefore defer a discussion of these until Sec.~\ref{sec:W_standardpaths}.

\section{Nontrivial Unitary Loops\label{sec:nontrivial_unitary_loops}}
\subsection{Properties of Loops}

While a loop evolution satisfies $U(1)=\id$ on a closed system, this will not generally be the case for the equivalent open system. To define the open system version of a unitary loop, we simply omit the terms in the generating Hamiltonian $H(t)$ that connect sites from either side of the boundary cut, formally writing
\be
H_{\rm op}(t)&=&H_{\rm cl}(t)-H_{\rm edge}(t)
\ee
for the terms in the Hamiltonian corresponding to the open system, closed system and sites across the boundary respectively. This is always a well-defined procedure, since a physical $H(t)$ may be written, at a given $t$, as a sum of local (and symmetry preserving, if required) terms. Similarly, we can define the open system version of a Floquet Hamiltonian by excluding the terms from $H_F$ that connect sites across the boundary.

Using Lieb-Robinson bounds \cite{Lieb:1972wy}, it can be shown that the open system unitary evolution corresponding a loop in $d$ dimensions differs from the identity only in a $(d-1)$-dimensional region close to the boundary (see argument in Ref.~\cite{Harper:2016us}). We call the nontrivial $(d-1)$-dimensional component of this open system evolution at the end of the drive the `effective edge unitary',
\beq
U_{\rm op}(1)=U_{\rm eff}.
\eeq
We will find that effective edge unitaries corresponding to nontrivial loops are anomalous, in that they may not be generated by any local (and symmetry-respecting) $(d-1)$-dimensional Hamiltonian.

The observation that running a loop on an open system leads to an effective unitary at the boundary leads us to propose a definition of equivalence among loops. We regard two loops acting on the same $d$-dimensional region $A$ as topologically equivalent if and only if the effective edge unitary of one can be obtained from the effective edge unitary of the other through the action of a $(d-1)$-dimensional symmetry-preserving unitary time evolution whose action is confined to the boundary $\delta A$. As presented, it might seem that this definition of equivalence depends on the choice of boundary $\delta A$. In Appendix~\ref{app:boundary_choice}, however, we show that this is not the case.

This notion of loop equivalence is connected to another, homotopic notion of loop equivalence. We show in Appendix~\ref{app:loop_equivalence} that if a loop $L_1$ can be continuously deformed to a loop $L_2$, then $L_1$ and $L_2$ are topologically equivalent in the sense defined above, i.e. the effective edge unitary of one can be obtained from the effective edge unitary of the other through a $(d-1)$ dimensional unitary transformation at the edge. It follows from this that two unitary evolutions which can be continuously deformed into each other have trivial relative order (as defined in Sec.~\ref{sec:rel_phase}). It also follows that it is impossible to find a continuous path connecting unitary evolutions which have a nontrivial relative topological order. This is the analog in driven systems of the statement that SPTs with different topological order cannot be connected through an adiabatic path.

\subsection{Types of Nontrivial Loop}
Having discussed some important properties of unitary loops, we now discuss the different types of nontrivial loop that are the main concern of this paper. As introduced previously, a nontrivial loop is one which cannot be contracted to the trivial loop ($U(t)=\id$) within the space $S_0$ or $S_G$. In two dimensions, the set of noncontractible loops in $S_0$ were discussed in Refs.~\cite{Harper:2016us,Po:2016uwb}, where it was found they form anomalous chiral phases that may be labeled by a pair of coprime integers.

\begin{figure}[t]
\begin{center}
\includegraphics[clip=true, trim = 0 0 0 0,scale=0.5]{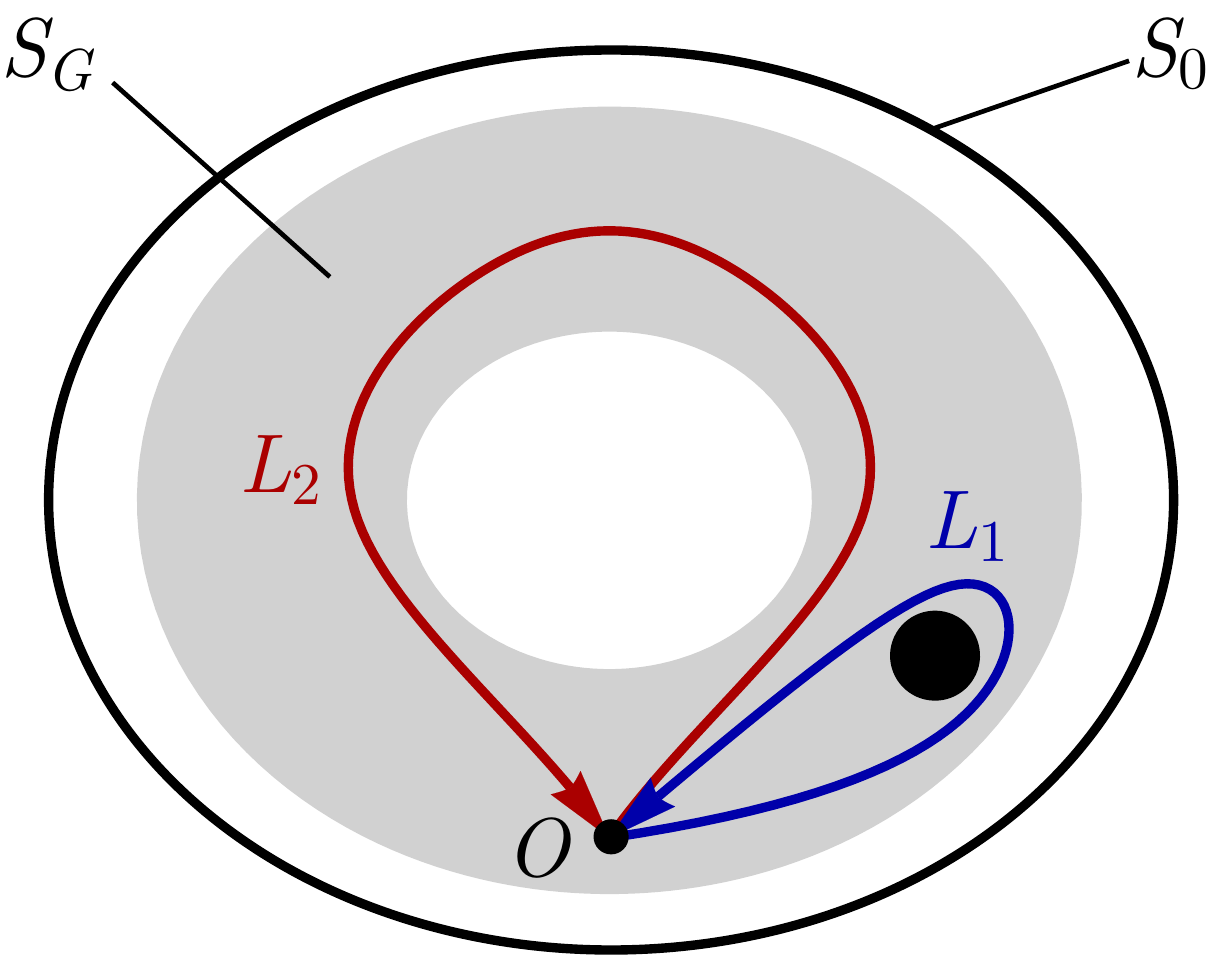}
\end{center}
\caption{Two distinct types of symmetry-preserving loops. The gray region corresponds to $S_G$, while the gray and white regions inside the large ellipse correspond to $S_0$. The filled black circle lies outside of both $S_0$ and $S_G$. $L_1$ shows a nontrivial FST loop that is noncontractible in both $S_G$ and $S_0$; $L_2$ shows a nontrivial FSPT loop that is noncontractible in $S_G$ but contractible in $S_0$.\label{fig:looptypes}}
\end{figure}

For our purposes, we are primarily concerned with noncontractible loops that are protected by some symmetry group $G$, which fall into two qualitatively different categories (see Fig.~\ref{fig:looptypes}). The first set consists of loops that are noncontractible in $S_G$ but which are also noncontractible in $S_0$---in other words, they lead to anomalous edge unitaries that cannot be generated by \emph{any} local Hamiltonian evolution at the boundary (even ignoring symmetry restrictions). These nontrivial loops are related to the anomalous chiral phases of Refs.~\cite{Harper:2016us,Po:2016uwb}, and we refer to them as Floquet  symmetric topological drives (FST drives).

The second set of symmetry-preserving noncontractible loops consist of loops that are noncontractible in $S_G$ but contractible in $S_0$. In other words, their effective edge unitaries may be generated by a local Hamiltonian at the boundary that breaks the protecting symmetry. We call these drives Floquet symmetry-protected topological drives (FSPT drives) in analogy with the nomenclature of static systems, because in the absence of symmetry, these loops are trivial. Under appropriate conditions, these FSPT drives may be used to generate FSPT phases.

In contrast to static systems, however, we can combine symmetry-protected topological drives from each class, either by running one nontrivial loop after the other, or by running two nontrivial loops concurrently on a tensor product of Hilbert spaces. For a sequence of two drives, it is clear that the spaces $S_0$ and $S_G$ are closed. However, an FST drive followed by an FSPT drive forms a hybrid loop in the space $S_G$, but is contractible to just the FST loop in the space $S_0$.

Similarly, the spaces $S_0$ and $S_G$ are also closed with respect to the loop stacking operation. If we stack an FST drive with an FSPT drive, then the complete evolution forms a hybrid loop in the space $S_G$, but is equivalent to just the FST drive (stacked with a trivial evolution) in the space $S_0$.

In the next section, Sec.~\ref{sec:TSDs}, we discuss FST drives and their relation to the previously studied anomalous chiral drives of Refs.~\cite{Po:2016uwb,Harper:2016us}. In Sec.~\ref{sec:FSPT_not_cohomology}, we discuss a type of symmetry-protected drive, based on these chiral drives, which are qualitatively distinct from other FSPT drives. In Sec.~\ref{sec:cohomology_2d} and Sec.~\ref{sec:cohomology_anyd}, the bulk of this paper considers FSPT drives from a group cohomology perspective, beginning with examples from 2d, before generalizing to all dimensions.
\section{Floquet Symmetric Topological Drives in Two Dimensions\label{sec:TSDs}}
In this section, we consider two-dimensional FST drives that correspond to nontrivial loops in $S_G$, but which are also noncontractible as loops in $S_0$. The existence of drives of this form was demonstrated in Refs.~\cite{Po:2016uwb,Harper:2016us}, and we now recall some of their features. 

An anomalous chiral drive in $S_0$ may be constructed from a set of elementary exchange moves, which act on the Hilbert spaces of two neighboring sites ($\mathcal{H}_\br\otimes\mathcal{H}_{\br'}$) with the action
\beq
U_{\br\br'}\ket{\br,\alpha}\otimes\ket{\br',\beta}&=&\ket{\br,\beta}\otimes\ket{\br',\alpha},
\eeq
where $\alpha$ and $\beta$ are the initial states on sites $\br$ and $\br'$ respectively. From these elementary moves, it is possible to construct drives which have an effective edge unitary that acts as a translation along the boundary.

To find examples of chiral anomalous drives in $S_G$, we need to impose the restrictions that arise from the symmetry group. To do this, we assume that the state on each site is labeled by an element $g_\alpha$ that belongs to the (left) regular representation of $G$. The global group action $V(g)$, corresponding to the group element $g$, then acts on each on-site group element by left multiplication. With this setup, it is simple to verify that a generic exchange move commutes with the action of the group, since 
\beq
V(g)U_{\br\br'}\ket{\br,g_\alpha}\otimes \ket{\br',g_\beta}&=&\ket{\br,gg_\beta}\otimes \ket{\br',gg_\alpha}\\
&=&U_{\br\br'}V(g)\ket{\br,g_\alpha}\otimes \ket{\br',g_\beta}.\nonumber
\eeq

Furthermore, if the group $G$ has a decomposition $G=A\times B$, then we may write the local group element at each site as $g_\alpha=(a_\alpha,b_\alpha)$, where $a_\alpha\in A$ and $b_\alpha\in B$. This allows us to consider exchanges that swap only one of the two components,
\begin{equation}
U^a_{\br\br'}\ket{\br,(a_\alpha,b_\alpha)}\otimes \ket{\br',(a_\beta,b_\beta)}=\ket{\br,(a_\beta,b_\alpha)}\otimes \ket{\br',(a_\alpha,b_\beta)}.
\end{equation}
As for the exchange swap above, this partial exchange operation also commutes with the group action, since
\beq
&&V(a)U^a_{\br\br'}\ket{\br,(a_\alpha,b_\alpha)}\otimes \ket{\br',(aa_\beta,b_\beta)}\nonumber\\
&=&\ket{\br,(aa_\beta,b_\alpha)}\otimes \ket{\br',(a_\alpha,b_\beta)}\\
&\equiv&U^a_{\br\br'}V(a)\ket{\br,(a_\alpha,b_\alpha)}\otimes \ket{\br',(a_\beta,b_\beta)},\nonumber
\eeq
and the commutation with $V(b)$ is trivial. By forming a four-step drive out of the pairwise exchange operations $U^a_{\br\br'}$ in the manner described in Ref.~\cite{Harper:2016us}, an effective edge unitary can be generated that uniformly translates group elements $a_\alpha$ around the boundary of an open system. This drive cannot be generated by any local Hamiltonian at the edge (symmetric or otherwise), and is noncontractible in both $S_0$ and $S_G$.

More generally, a group $G$ may have two different decompositions, $A_1\times B_1$ and $A_2\times B_2$, where the $A_i$ are the same size and the $B_i$ are also the same size. A topological symmetric drive that translates the component $A_1$ and a similar drive that translates the component $A_2$ would be equivalent in $S_0$, but are not necessarily equivalent in $S_G$. We leave a discussion of these ideas, and a consideration of groups which do not have a direct product decomposition, to future work.

\section{FSPT Drives Outside of the Group Cohomology Construction\label{sec:FSPT_not_cohomology}}
Motivated by the topological symmetric drives discussed above in Sec.~\ref{sec:TSDs}, we might also consider a related set of FSPT drives (i.e. trivial in $S_0$) that are based on a similar construction, and which are qualitatively different from the group cohomological FSPT drives we introduce in the next section. 

Consider a group with decomposition $G=A\times B$, in which the groups $A$ and $B$ have the same size (and which may correspond to the same group). We may now construct a drive which, at the boundary, translates the on-site $a$-component to the left but the on-site $b$ component to the right. In $S_0$, this drive transfers no net information, and may be contracted to the identity. In $S_G$, however, this loop is not necessarily contractible. The group $G={\rm U(1)}\times {\rm U(1)}$ gives an example of this, as we now demonstrate.

We consider two species of hardcore boson on a two-dimensional lattice, where particle number is conserved separately for each species. The on-site Hilbert space at site $\br$ may be represented by two numbers, $(n^a_\br, n^b_\br)$, where $n^j_\br\in\{0,1\}$ represents the number of particles of species $j$ at site $\br$.

\begin{figure}[t]
\begin{center}
\includegraphics[clip=true, trim = 0 0 0 0,scale=0.5]{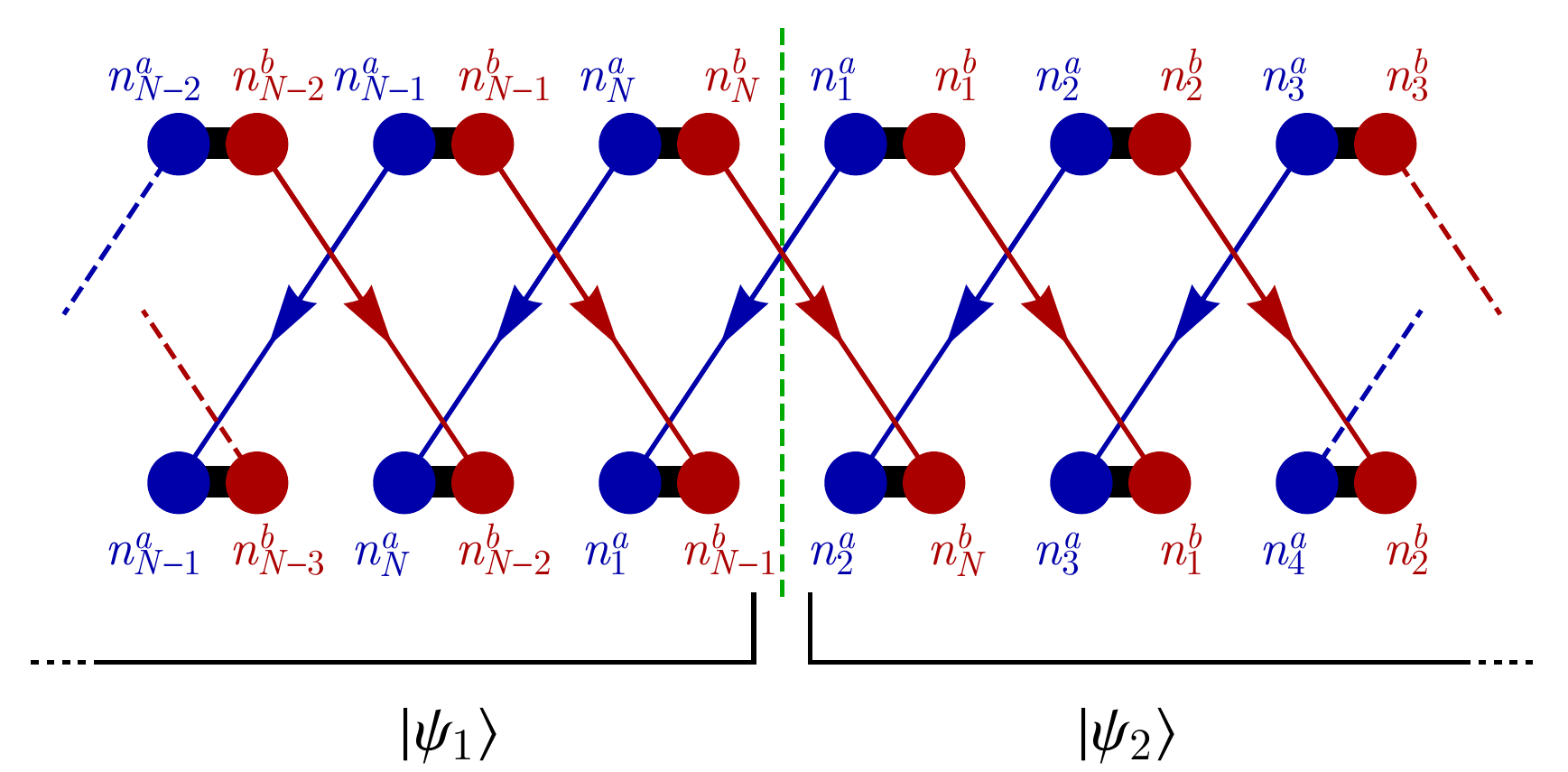}
\end{center}
\caption{Edge action of a nontrivial FSPT drive that lies outside of the group cohomology construction. Blue $a$-particles are translated to the left, while red $b$-particles are translated to the right. The dashed green vertical line indicates where the chain should be cut to form the open edge described in the main text. The states $\ket{\psi_1}$ and $\ket{\psi_2}$ include $M$ sites either side of the cut, and are acted upon by the effective edge unitary $U_{\rm edge}$. See main text for details. \label{fig:FSPTedge}}
\end{figure}

We may now construct a model whose effective edge unitary translates the occupation of the $a$-particles to the left and the $b$-particles to the right along the boundary. Explicitly, this drive can be constructed by stacking two copies of the model of Ref.~\cite{Harper:2016us}, one for each species of particle. Labeling the sites at the boundary from $j=1$ to $j=N$, the effective edge unitary has the action
\beq
&&U_{\rm eff}\ket{(n^a_{1},n^b_{1});\ldots (n^a_{i},n^b_{i});\ldots(n^a_{N},n^b_{N})}\\
&=&\ket{(n^a_{2},n^b_{N});\ldots (n^a_{i+1},n^b_{i-1});\ldots(n^a_{1},n^b_{N-1}},\nonumber
\eeq
as shown in Fig.~\ref{fig:FSPTedge}. Since the amount of quantum information transported for each species is equal and opposite, this is a trivial drive in $S_0$ \cite{Harper:2016us}. However, on the basis of charge conservation arguments, as long as the two species separately have number conservation, we are led to argue that this is a nontrivial loop in $S_G$. This is equivalent to the statement that the effective edge unitary cannot be generated by a local, symmetric unitary evolution at the boundary.

We will show this by contradiction, and therefore assume initially that $U_{\rm eff}$ \emph{can} be generated by a local, one-dimensional, symmetric Hamiltonian, $H_{\rm eff}$. If this is the case, then we can also define unitary evolutions for an open boundary (which has been cut into a one-dimensional open chain), by excluding the terms in $H_{\rm eff}$ that connect sites across the cut. Due to Lieb-Robinson bounds \cite{Lieb:1972wy}, following the arguments presented in Ref.~\cite{Harper:2016us}, it follows that $U_{\rm eff}^{\rm open}=U_{\rm edge}U_{\rm eff}^{\rm closed}$, where $U_{\rm eff}^{\rm open/closed}$ is the unitary evolution for the closed/open boundary, and $U_{\rm edge}$ is a local unitary evolution which has a finite region of influence near each boundary. With this interpretation, $U_{\rm edge}$ may be seen as a unitary transformation on this space that takes a state of the final evolved closed system to a state of the final evolved open system.

Specifically, we assume that the action of $U_{\rm edge}$ is confined to a $2^{2M}$-dimensional Hilbert space (spanning $M$ sites either side of the cut). For a given initial state, 
\beq
\ket{(n^a_{1},n^b_{1})\ldots (n^a_{i},n^b_{i});\ldots(n^a_{N},n^b_{N})},
\eeq
the state in the space that $U_{\rm edge}$ acts on is $\ket{\psi_1}\otimes\ket{\psi_2}$, where
\beq
\ket{\psi_1}&=&\ket{(n^a_{N-M+2},n^b_{N-M})\ldots,(n^a_{1},n^b_{N-1})}\\
\ket{\psi_2}&=&\ket{(n^a_2,n^b_{N});(n^a_3,n^b_1);\ldots,(n^a_{M+1},n^b_{M-1})},
\eeq
as shown in Fig.~\ref{fig:FSPTedge}. Through a unitary transformation, we change basis so that this state may be written,
\beq
\ket{\tilde{\psi}_1}&=&\ket{(n^a_{N-M+2},n^b_{N-M})\ldots,(n^b_{N},n^b_{N-1})}\\
\ket{\tilde{\psi}_2}&=&\ket{(n^a_2,n^a_{1});(n^a_3,n^b_1);\ldots,(n^a_{M+1},n^b_{M-1})}.
\eeq
Acting in this new basis, $U_{\rm edge}$ must decompose into a product form, $U_L\otimes U_R$, where $U_L$ acts on $\ket{\tilde{\psi}_2}$ and produces a state at the left edge and $U_R$ acts on $\ket{\tilde{\psi}_1}$ to produce a state at the right edge. In addition, $U_L$ and $U_R$ must individually preserve the $U(1) \times U(1)$ symmetry. If the unitary did not have this form, the state which results from the action of $U_{\rm eff}^{\rm open}$ would have an entanglement of the bits at the left and right edges, which would violate the Lieb-Robinson bound on the propagation of information.

Now, $U_L$ maps the state $\ket{\tilde{\psi}_2}$ onto some linear combination of final states of the form 
\beq
\ket{(n^{a\prime}_1,n^{b\prime}_1);(n^{a\prime}_2,n^{b\prime}_2);\ldots(n^{a\prime}_M,n^{b\prime}_M)}
\eeq
(where the $n^{j\prime}_i$ are unknown), which must hold for an arbitrary initial configuration of occupation numbers in $\ket{\tilde{\psi}_2}$. However, the state in which all of the initial $n^a$ occupation factors are one has total $a$-charge $M+1$, and cannot be mapped onto any possible final state, which has maximum $a$-charge of $M$. This is a contradiction, and so this drive cannot be generated by a local, symmetry-preserving unitary evolution at the boundary. 

The above construction can also be extended to other continuous groups. For instance, by taking $G=SU(2)\times SU(2)$ and labeling on-site states by the eigenvalues of the total spin, we can construct drives which translate the two different spin species in different directions, resulting in a drive that is trivial in $S_0$ but nontrivial in $S_G$. These considerations also apply to a large class of other groups, including products of groups of the form above. In addition, the elementary drives may be stacked and run in sequence to give more complicated drives. For the groups ${\rm U(1)\times U(1)}$, the set of drives obtained in this way has a natural $\zbb$ classification (from the considerations of Ref.~\cite{Po:2016uwb,Harper:2016us}).

We note, however, that these arguments do not apply to finite groups. It is possible to show, at least for some specific cases such as $\zbb_2\times\zbb_2$, that similarly constructed drives are topologically trivial within $S_G$. This is in contrast to the group cohomology construction that we introduce in the next section, which requires Hilbert spaces that scale as the size of the group, and so works for finite groups but fails for continuous groups.

\section{Group Cohomology Construction of FSPT Drives in Two Dimensions\label{sec:cohomology_2d}}
We now introduce a set of FSPT drives based on  group cohomology elements. These drives are noncontractible unitary loops that act as the identity in a closed system, but which generate SPT order at the boundary of an open system. In this section, we consider two-dimensional drives on a square lattice, where the underlying concepts can be seen most intuitively, and we will use the group $\zbb_2\times\zbb_2$ as an illustrative example. In Sec.~\ref{sec:cohomology_anyd}, we generalize the models to higher dimensions and more general lattices and groups.
\subsection{$\zbb_2\times\zbb_2$ Model in 2d}
\begin{figure}[t]
\begin{center}
\includegraphics[scale=0.6]{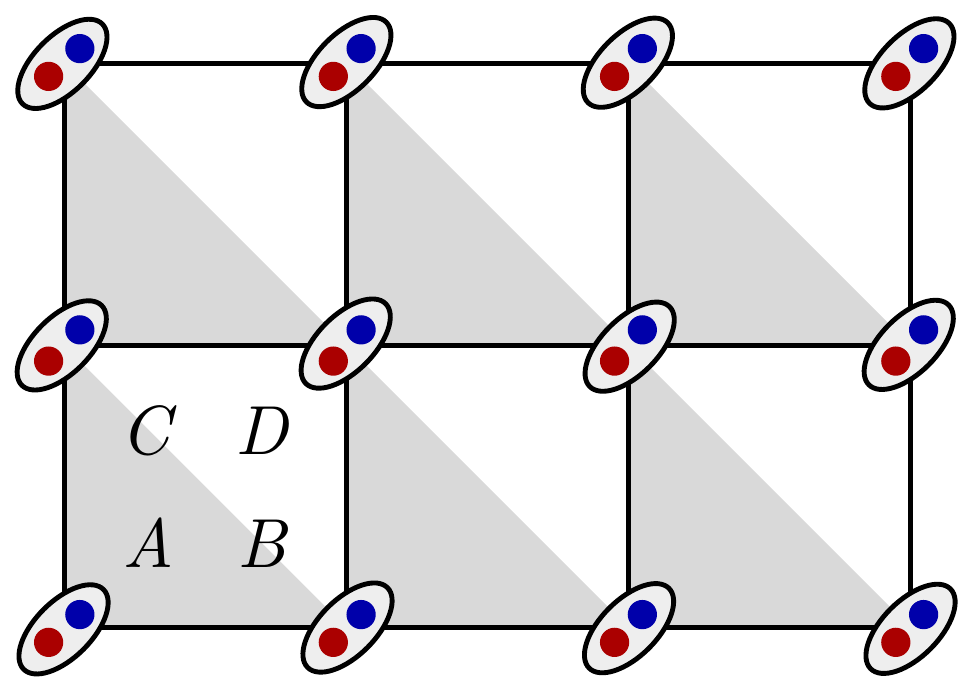}
\end{center}
\caption{Square lattice for the 2d $\zbb_2\times\zbb_2$ model. Red and blue points represent the two $\zbb_2$ degrees of freedom. Letters $ABCD$ label the four sites of a given plaquette. Shading indicates the separation of each plaquette into two triangles. \label{fig:z2z2model}}
\end{figure}

As a motivating example, we first describe a specific two-dimensional model protected by the symmetry $G=\zbb_2\times\zbb_2$. We consider a square lattice ($\Lambda$) with two $\zbb_2$ degrees of freedom per site, represented by points in Fig.~\ref{fig:z2z2model}. 
The two generators of the $\zbb_2\times\zbb_2$ symmetry may be written
\beq
V(\sigma)&=&\prod_{\br\in\Lambda}\sigma^x_{\br}\label{eq:z2z2symgen}\\
V(\tau)&=&\prod_{\br\in\Lambda}\tau^x_{\br},\nonumber
\eeq
where $\sigma$ and $\tau$ are Pauli matrices, and we can label a general state by its eigenvalues under the on-site operators $\{\sigma^z_\br,\tau^z_\br\}$, which we write as $\{s_\br,t_\br\}$.

To define the unitary drive, it is helpful to imagine each square plaquette as being formed of two triangles, indicated by shading in Fig.~\ref{fig:z2z2model}. We also label the sites within each plaquette with the letters $ABCD$. Consider the following two-site Hamiltonian, which acts, for example, on the $A$ and $B$ sites in a single plaquette,
\beq
\h_{AB}^{(2)}&=&\frac{\pi}{8}\left[\sigma^z_A\sigma^z_B\tau^z_B+\sigma^z_B\tau^z_A+\sigma^z_A\tau^z_A\tau^z_B\right]\\
&&-\frac{\pi}{8}\left[\sigma^z_A\sigma^z_B\tau^z_A+\sigma^z_A\tau^z_B+\sigma^z_B\tau^z_A\tau^z_B\right].\nonumber
\eeq
It is clear that $\h^{(2)}_{AB}$ does not preserve the global symmetry $G$, since it does not commute with the symmetry generators in Eq.~\eqref{eq:z2z2symgen}. However, if we evolve with this Hamiltonian until time $t=1$, then the endpoint unitary satisfies the following simple relations under the action of the symmetry,
\beq
V(\sigma)e^{-i\h_{AB}^{(2)}}V(\sigma)^{-1}&=&e^{-i\h_{AB}^{(2)}-i\h_{B}^{\sigma}+i\h_{A}^{\sigma}}\label{eq:z2z2ham2sym}\\
V(\tau)e^{-i\h_{AB}^{(2)}}V(\tau)^{-1}&=&e^{-i\h_{AB}^{(2)}-i\h_{B}^{\tau}+i\h_{A}^{\tau}},\nonumber
\eeq
where the single-site Hamiltonians are
\beq
\h^\sigma_{A}&=&-\frac{\pi}{4}\sigma_A^z+\frac{\pi}{4}\sigma_A^z\tau_A^z\label{eq:z2z2ham1}\\
\h^\tau_{A}&=&\frac{\pi}{4}\tau_A^z-\frac{\pi}{4}\sigma_A^z\tau_A^z.\nonumber
\eeq

This motivates a three-site Hamiltonian, $\h_{ABC}^{(2)}=\h^{(2)}_{AB}+\h^{(2)}_{BC}+\h^{(2)}_{CA}$, which cyclically connects the sites around a shaded triangle in Fig.~\ref{fig:z2z2model}. If we evolve the system until $t=1$ with this new Hamiltonian, then the final unitary $e^{-i\h_{ABC}^{(2)}}$ \emph{will} be symmetric under the symmetry generators, since the nonsymmetric single-particle components in Eq.~\eqref{eq:z2z2ham2sym} will cancel out. We may formally take the logarithm of this unitary operator to find the effective three-site Hamiltonian
\beq
\h_{ABC}^{(3)}&=&\frac{\pi}{8}\bigg[\sigma_A^z\sigma_C^z\tau_A^z\tau_B^z+\sigma_B^z\sigma_C^z\tau_A^z\tau_C^z+\sigma_A^z\sigma_B^z\tau_B^z\tau_C^z\\
&&-\left(\sigma_B^z\sigma_C^z\tau_A^z\tau_B^z+\sigma_A^z\sigma_B^z\tau_A^z\tau_C^z+\sigma_A^z\sigma_C^z\tau_B^z\tau_C^z\right)\bigg],\nonumber
\eeq
which is now explicitly symmetric under $V(\sigma)$ and $V(\tau)$ (since each term contains an even number of $\sigma^z$ and $\tau^z$ operators).

We extend this idea to the complete lattice, defining the global Hamiltonian
\beq
\h_{\zbb_2\times\zbb_2}&=&\sum_{\rm plaquettes}\left[\h_{ABC}^{(3)}-\h_{BCD}^{(3)}\right],\label{eq:hz2z2}
\eeq
which is a sum of symmetric, commuting terms (note that we choose a different orientation for shaded and unshaded triangles). From the properties above, this generates a unitary evolution that may be written at time $t=1$ as
\beq
e^{-i\h_{\zbb_2\times\zbb_2}}&=&\prod_{\rm plaquettes}e^{-i\h^{(2)}_{AB}-i\h^{(2)}_{CA}+i\h^{(2)}_{CD}+i\h^{(2)}_{DB}},
\eeq
where two terms in the exponent proportional to $\h_{BC}^{(2)}$ have cancelled out. Moreover, when the product over plaquettes is taken, most of the terms in the exponent will again cancel, since every pair of neighboring sites is shared by two plaquettes, which generate two-site terms in the exponent with opposite signs. The only nonzero terms will arise at the edges of the system where there is no cancellation; for a closed system, the unitary operator is the identity.

The Hamiltonian $\h_{\zbb_2\times\zbb_2}$ therefore generates a unitary loop. When evolved until $t=1$, the evolution acts as the identity in the bulk, but leads to an effective edge unitary at the boundary of an open system. In fact, as argued in the next section, this unitary drive effectively converts a product state at the edge to a non-trivial 1d SPT state, and so the drive is nontrivial and describes a dynamical 2d SPT phase. The novel behavior of this unitary evolution is related to the nontrivial projective representation of the group $\zbb_2\times\zbb_2$. We now extend this type of drive to other groups.
\subsection{Projective Representations\label{sec:projrep}}
In general, we aim to produce a nontrivial unitary drive using a Hamiltonian $H_G$ that is protected by some symmetry group $G$. The drive should behave trivially in the bulk, but should lead to SPT order at the 1d boundary of an open system. Since 1d SPTs are in one-to-one correspondence with the projective representations of the group $G$ \cite{Pollmann:2010ih,Schuch:2011ip,Chen:2011iqa,Pollmann:2010iha}, we begin by recalling some properties of projective representations. 

We write elements of the group $G$ as $g_i$, and recall that operators $P_\omega(g)$ form a projective representation of $G$ if and only if
\beq
P_\omega(g_a)P_\omega(g_b)&=&\omega(g_a,g_b)P_\omega(g_ag_b)
\eeq
for all $g_a,g_b\in G$, where $\omega(g_a,g_b)\in U(1)$ belongs to the factor system of the projective representation. If all $\omega(g_a,g_b)$ are trivial, then the $P_\omega(g)$ form a linear representation. More generally, the factor system has the property
\beq
\omega(g_a,g_b)\omega(g_ag_b,g_c)&=&\omega(g_b,g_c)\omega(g_a,g_bg_c).\label{eq:structure_factor}
\eeq
For the current purpose, it is useful to define a related set of U(1) phases through
\beq
\alpha(g_a, g_ag_b)&=&\omega(g_a,g_b).\label{eq:alphadef}
\eeq
(Note that both of these phases are related to 2-cocycles from the theory of group cohomology, a relationship we make explicit in Sec.~\ref{sec:cohomology_anyd}). In terms of the $\alpha$ phases, the relation in Eq.~\eqref{eq:structure_factor} may be written in the useful form
\beq
\alpha(g_a,g_b)&=&\alpha(gg_a,gg_b)\frac{\alpha(g,gg_a)}{\alpha(g,gg_b)}.\label{eq:alpha_relation}
\eeq
In Appendix~\ref{app:z2z2}, we explicitly consider the nontrivial projective representation of the Abelian group $G=\zbb_2\times\zbb_2$, giving the U(1) phases $\omega(g_a,g_b)$ and $\alpha(g_a,g_b)$.
\subsection{Nontrivial Unitary Loop}
We set up the system on a square lattice, where the state on each site is an element from the (left) regular representation of $G$. Each site can therefore be labeled by a group element, which we write (for site $\br$) as $g_\br$.


As for the $\zbb_2\times\zbb_2$ model, we label the four sites within each plaquette as $ABCD$ and interpret each square plaquette as being formed from two triangles, as in Fig.~\ref{fig:z2z2model}. The Hamiltonian is then a sum over all group element configurations on each plaquette,
\beq
\h_{G}&=&-\sum_{\rm plaquettes}\sum_{g_A,g_B,g_C,g_D}\bigg[\\
&&H_{\blacktriangle}(g_A,g_B,g_C)\ket{g_A,g_B,g_C}\bra{g_A,g_B,g_C}\nonumber\\
&&+H_{\triangledown}(g_B,g_C,g_D)\ket{g_B,g_C,g_D}\bra{g_B,g_C,g_D}\bigg],\nonumber
\eeq
with coefficients
\begin{align}
H_{\blacktriangle}(g_A,g_B,g_C)&=\mathrm{Arg}\left[\alpha\left(g_A,g_B\right)\alpha\left(g_B,g_C\right)\alpha\left(g_C,g_A\right)\right]\\
H_{\triangledown}(g_B,g_C,g_D)&=-\mathrm{Arg}\left[\alpha\left(g_B,g_C\right)\alpha\left(g_C,g_D\right)\alpha\left(g_D,g_B\right)\right],
\end{align}
and where the Arg function takes values in the range $(-\pi,\pi]$. It is implied that each term in the Hamiltonian acts as the identity everywhere outside of the three sites it depends upon explicitly, and it may be verified that each term in the Hamiltonian commutes with every other. When $G=\zbb_2\times\zbb_2$, $H_G$ may be reduced to the expression for $H_{\zbb_2\times\zbb_2}$ given in Eq.~\eqref{eq:hz2z2}.

We note a number of useful properties of this Hamiltonian. First, it preserves the symmetry associated with the group $G$. To see this, we act with $V(g)$, the global symmetry operator associated with the regular representation of the group element $g$, to find
\beq
&&V(g)\h_{G}V(g)^{-1}=\sum_{\rm plaquettes}\sum_{g_A,g_B,g_C,g_D}\bigg[\\
&&H_{\blacktriangle}(g_A,g_B,g_C)\ket{gg_A,gg_B,gg_C}\bra{gg_A,gg_B,gg_C}\nonumber\\
&&+H_{\triangledown}(g_B,g_C,g_D)\ket{gg_B,gg_C,gg_D}\bra{gg_B,gg_C,gg_D}\bigg].\nonumber
\eeq
However, using Eq.~\eqref{eq:alpha_relation}, we see that
\beq
&&\alpha\left(g_A,g_B\right)\alpha\left(g_B,g_C\right)\alpha\left(g_C,g_A\right)\\
&=&\alpha\left(gg_A,gg_B\right)\alpha\left(gg_B,gg_C\right)\alpha\left(gg_C,gg_A\right),\nonumber
\eeq
and so $H_{\blacktriangle}(g_A,g_B,g_C)=H_{\blacktriangle}(gg_A,gg_B,gg_C)$ and $H_{\triangledown}(g_B,g_C,g_D)=H_{\triangledown}(gg_B,gg_C,gg_D)$. Since the Hamiltonian sums over all configurations of group elements, it follows that
\beq
V(g)\h_{G}V(g)^{-1}&=&\h_{G},
\eeq
and the Hamiltonian is symmetric under $G$.

\begin{figure}[t!]
\begin{center}
\includegraphics[scale=0.5]{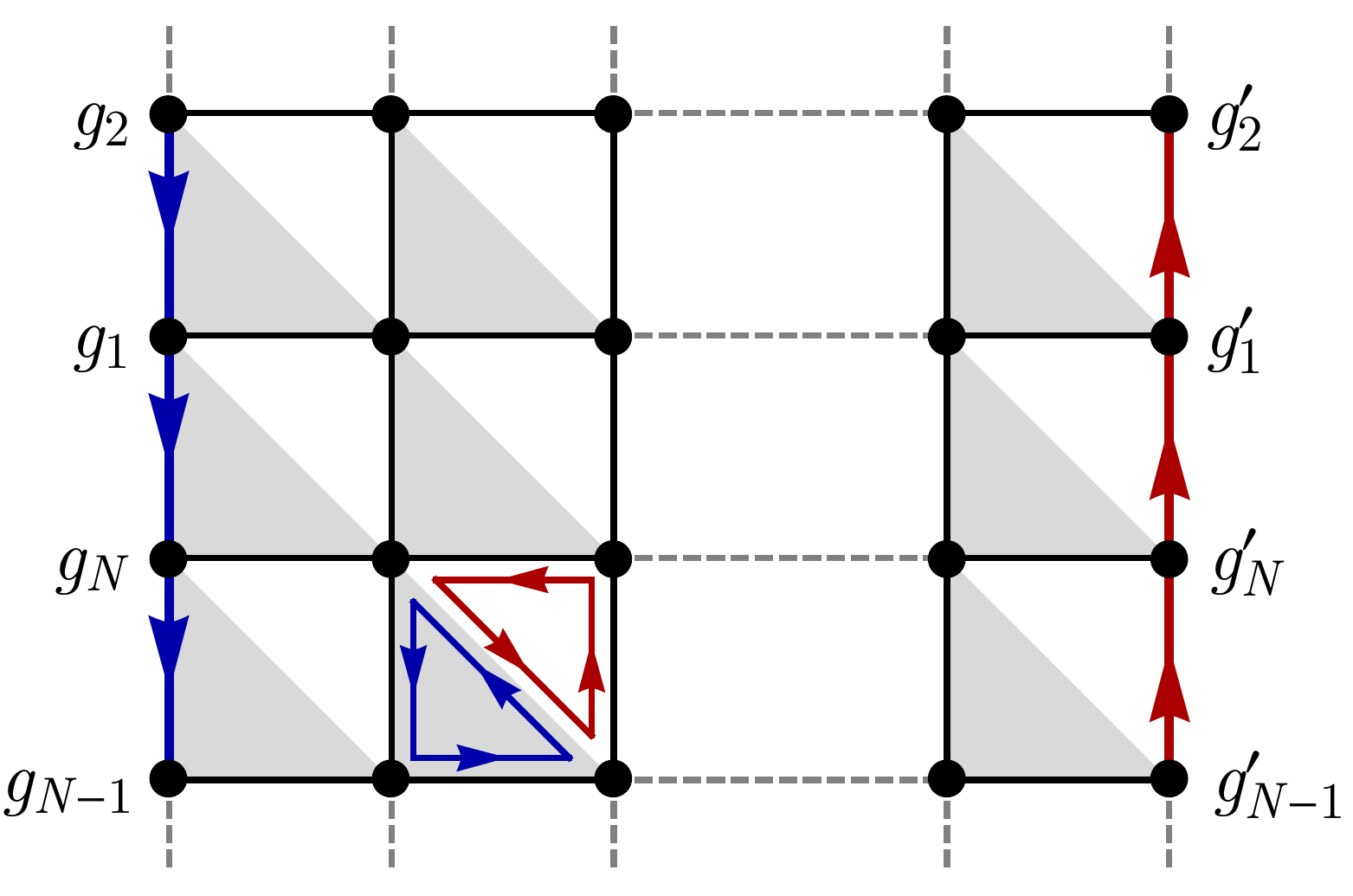}
\end{center}
\caption{Action of $U_G(1)$ on a cylinder (for a general symmetry group $G$). States at the edge are labeled by a group element $g_j$ and sites at the edge are labeled from one to $N$. Periodic boundary conditions are in the vertical direction. Red and blue arrows indicate the cancellation of U(1) phases in the bulk and the non-cancellation of U(1) phases at the two edges. \label{fig:cylinder}}
\end{figure}

Next we consider the action of the unitary evolution, $U_{G}(1)=\exp\left(-i\h_{G}\right)$. Since $\h_{G}$ is a sum of commuting terms, the unitary evolution operator may be written
\begin{widetext}
\beq
U_{G}(1)&=&\prod_{\rm plaquettes}\prod_{g_A,g_B,g_C,g_D}\left[\overline{\ket{g_A,g_B,g_C}}\overline{\bra{g_A,g_B,g_C}}+e^{iH_{\blacktriangle}(g_A,g_B,g_C)}\ket{g_A,g_B,g_C}\bra{g_A,g_B,g_C}\right]\times\\
&&\left[\overline{\ket{g_B,g_C,g_D}}\overline{\bra{g_B,g_C,g_D}}+e^{iH_{\triangledown}(g_B,g_C,g_D)}\ket{g_B,g_C,g_D}\bra{g_B,g_C,g_D}\right],\nonumber
\eeq
\end{widetext}

where $\overline{\ket{g_A,g_B,g_C}}\overline{\bra{g_A,g_B,g_C}}$ projects onto the complement of $\ket{g_A,g_B,g_C}$ and where it may be seen that the factors

\beq
e^{iH_{\blacktriangle}(g_A,g_B,g_C)}&=&\alpha\left(g_A,g_B\right)\alpha\left(g_B,g_C\right)\alpha\left(g_C,g_A\right)\label{eq:2d_phasefactors}\\
e^{iH_{\triangledown}(g_B,g_C,g_D)}&=&\left[\alpha\left(g_B,g_C\right)\alpha\left(g_C,g_D\right)\alpha\left(g_D,g_B\right)\right]^{-1}.\nonumber
\eeq
Consider the action of $U_{G}(1)$ in the bulk. It is clear from Fig.~\ref{fig:z2z2model} that every pair of neighboring sites (including diagonal B--C neighbors) is affected by two factors in $U_{G}(1)$, one contributing a phase of the form $e^{iH_\blacktriangle}$ and the other contributing a phase of the form $e^{iH_\triangledown}$. From Eq.~\eqref{eq:2d_phasefactors}, the action of $U_{G}(1)$ due to the pair of neighboring states $g_\br$ and $g_{\br}$ is then
\bequ
U_{G}(1)\ket{\ldots g_\br,g_{\br'}\ldots}=\ldots\frac{\alpha(g_\br,g_{\br'})}{\alpha(g_{\br},g_{\br'})}\ldots\ket{\ldots g_\br,g_{\br'}\ldots},
\eequ
which is trivial. A similar cancellation occurs for every pair of neighboring sites in the bulk and so the operator $U_{G}(1)$ is the identity for a closed system. At the edges of an open system, however, its action is nontrivial, as we now demonstrate. Taking a cylinder with circumference $N$ and boundary states $\ket{\{g_i\}}\otimes\ket{\{g'_i\}}$, the action of $U_{G}(1)$ restricted to the two boundaries is
\beq
U_{G}(1)\ket{\{g_i\}}\otimes\ket{\{g'_i\}}&=&\frac{\alpha(g_2,g_1)\alpha(g_3,g_2)\ldots\alpha(g_1,g_N)}{\alpha(g_2',g_1')\alpha(g_3',g_2')\ldots\alpha(g_1',g_N')}\times\nonumber\\
&&\ket{\{g_i\}}\otimes\ket{\{g'_i\}},
\eeq
where we have labeled the sites consistently from one to $N$ on each edge of the cylinder (see Fig.~\ref{fig:cylinder}). Since each edge forms a connected loop, if we act with the symmetry generator $V(g)$ then Eq.~\eqref{eq:alpha_relation} shows that the action of the unitary will be unchanged.

By comparing the form of this resultant edge state to the form of a nontrivial 1d SPT state \cite{Chen:2013foa}, we see that the drive acts on a trivial symmetric product state,
\beq
\ket{\psi}&=&\bigotimes_{i} \left[\sum_{g_i}\ket{g_i}\right] \bigotimes_{i'}\left[\sum_{g'_i}\ket{g'_i}\right],
\eeq 
to produce a nontrivial SPT state at the edge.

\section{Group Cohomology Construction of FSPT Drives in Any Dimension\label{sec:cohomology_anyd}}
We now place the drives introduced in the previous section onto a more formal footing and, in the process, generalize to all dimensions and lattices. Our aim will again be to produce a unitary evolution that:
\begin{enumerate}
\item	is symmetric under some group $G$
\item	in a closed system yields $U(1)=\id$
\item	in an open system yields $U(1)=U_{\rm eff}$, a nontrivial effective edge unitary that generates a $(d-1)$-dimensional SPT phase from a trivial product state.
\end{enumerate}
The setup of these drives requires some mathematical background, which we begin by introducing.
\subsection{Simplicial Complexes, Branching Structures, and Group Cohomology}
In an effort to keep the discussion concise, we will describe the mathematical framework only briefly in this section. For further details, we refer the reader to Refs.~\cite{Brown:1614758,Chen:2013foa}.

We define the system on an oriented manifold, $M$, which is triangulated to form a simplicial complex. In most physical systems, the sites of a real $d$-dimensional lattice can be identified with the vertices of the simplicial complex, with higher dimensional simplices defined that connect nearby points. In general, there will be many different ways of triangulating the manifold $M$.  While the Hamiltonian we write down will depend on this choice, the topological order of the unitary evolution will not.

\begin{figure}[t]
\begin{center}
\includegraphics[scale=0.3]{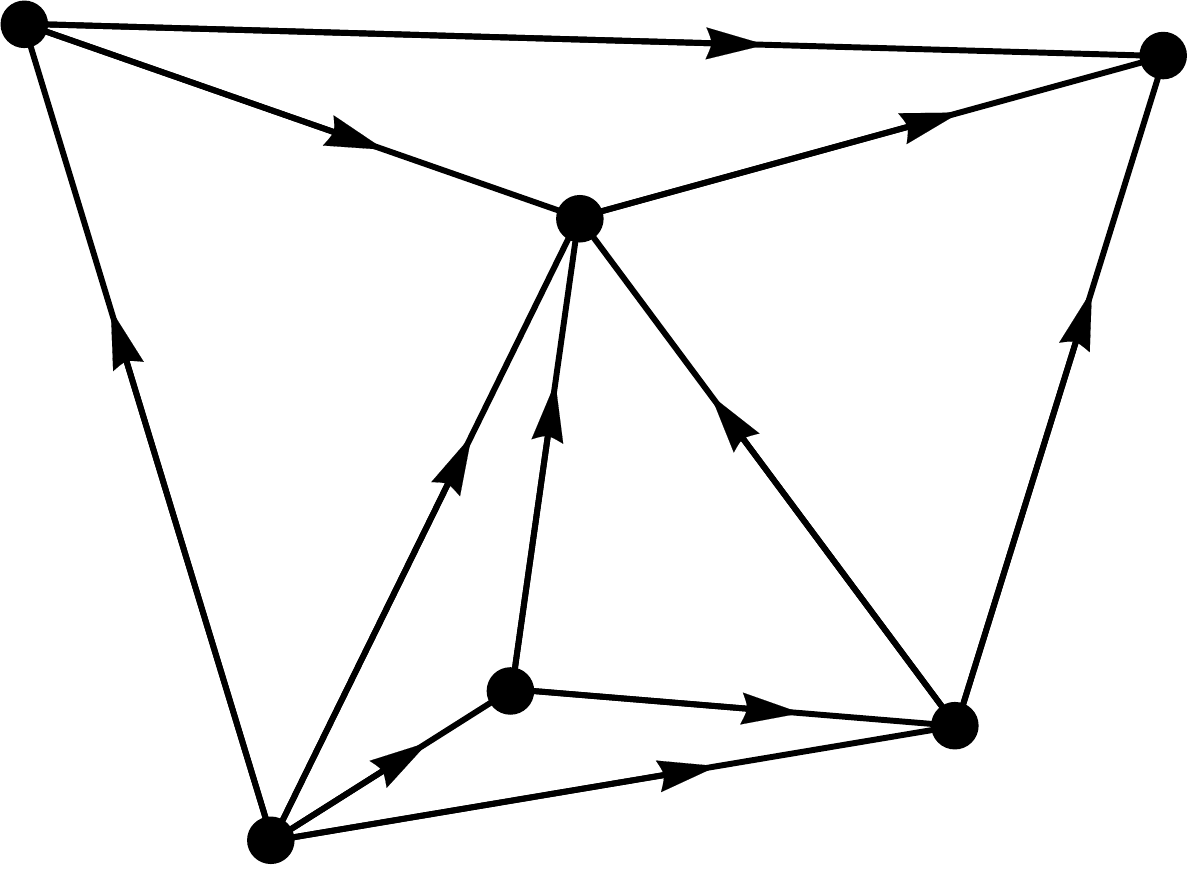}\hspace{1mm}
\includegraphics[scale=0.3]{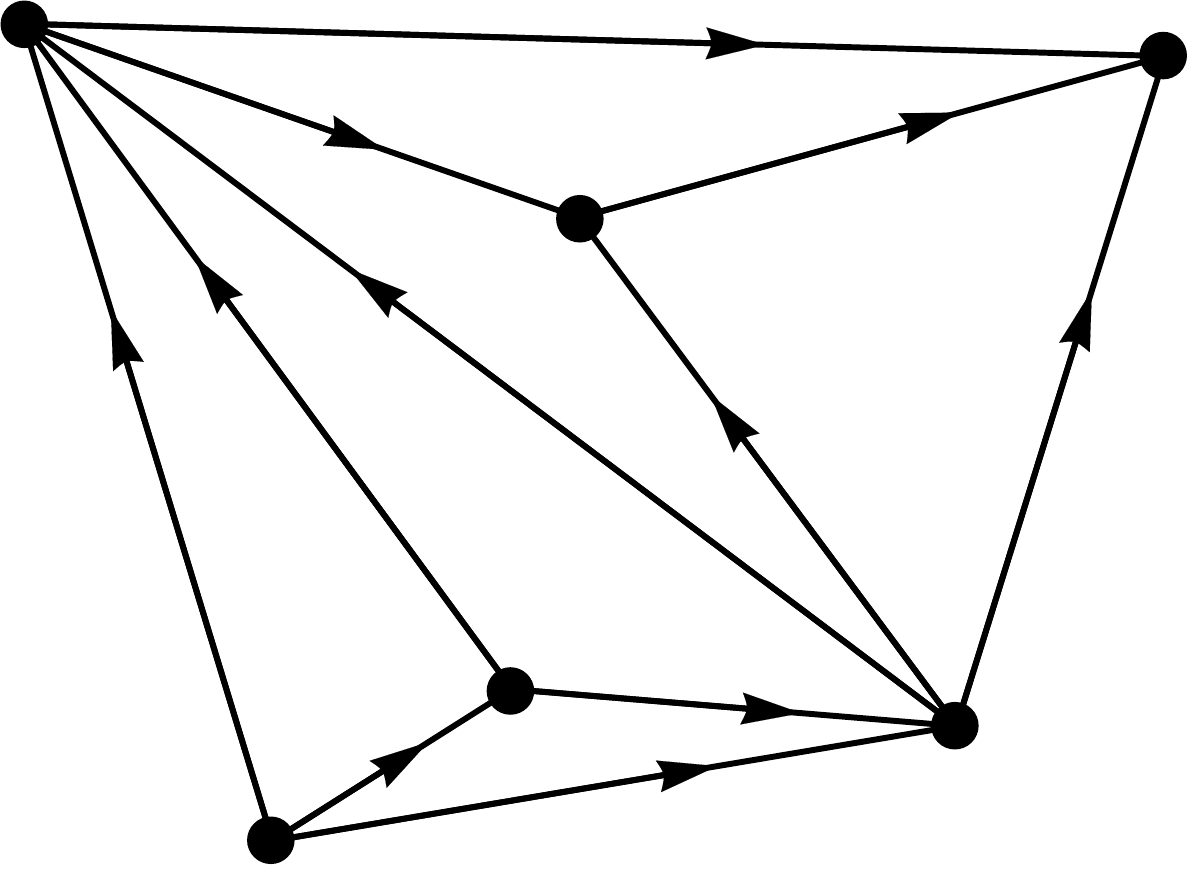}
\end{center}
\caption{Two possible triangulations and branching structures for a 2d set of points.\label{fig:tribranch}}
\end{figure}

We further equip  this oriented, simplicial complex with a branching structure \cite{Chen:2013foa}, which is a way of consistently assigning arrows to each line segment, such that there are no oriented loops around any single triangle. A simple way to generate a branching structure is to number the vertices of the simplicial complex from one to $N_V$, and then to draw an arrow on each edge in the direction of increasing index. There are many different possible branching structures for a given simplicial complex, but the topological order of the drives we define below are again independent of this choice. See Fig.~\ref{fig:tribranch} for two possible triangulations and branching structures for a 2d set of points.

The final piece of mathematical machinery we require is the notion of a nontrivial $n$-cocycle from the theory of group cohomology \cite{Brown:1614758,Chen:2013foa}. For a group $G$ and module $M$, an $n$-cochain is a map $\nu_n:G^{n+1}\to M$ that satisfies
\beq
g\cdot\nu_n(g_0,g_1,\ldots,g_n)\equiv \nu_n(gg_0,gg_1,\ldots,gg_n).\label{eq:cochainrel}
\eeq
For our purposes, the module $M$ is simply the Abelian group U(1). The set of cochains forms an Abelian group, $\mathcal{C}^n[G,M]$, under the multiplicative operation that takes
\bequ
\nu(g_0,g_1,\ldots,g_n)\nu'(g_0,g_1,\ldots,g_n)=\nu''(g_0,g_1,\ldots,g_n).
\eequ

In addition to the above, an operator $d_n$ may be defined that maps elements from $\nu_n$ onto elements from $\nu_{n+1}$ (see Ref.~\onlinecite{Chen:2013foa} for the explicit form of this operator). This allows us to define two useful Abelian subgroups of $\mathcal{C}^n[G,M]$: $\mathcal{Z}_n[G,M]$ is the set $n$-cocycles, which are elements of $\nu_n$ that satisfy $d_n\nu_n=0$, while $\mathcal{B}_n[G,M]$ is the set of $n$-coboundaries, which are elements of $\nu_n$ that take the form $\nu_n=d_{n-1}\nu_{n-1}$ for some $\nu_{n-1}$. The $n$th cohomology group is then given by ${H}^n[G,M]=\mathcal{Z}^n[G,M]/\mathcal{B}^n[G,M]$. In $d$ dimensions, we require elements from the $d$th cohomology group ${H}^d[G,M]$.

A nontrivial 2-cocycle, defined in this way, is in one-to-one correspondence with the factor system of a projective representation of $G$, through
\beq
\omega(g_a,g_b)&=&\nu_2(1,g_a,g_ag_b).
\eeq
If this substitution is made in the discussion below, then the 2d Hamiltonian introduced in the previous section can be recovered. 
\subsection{Group Cohomology Hamiltonian}
\begin{figure}[t]
\begin{center}
\includegraphics[clip=true, trim = 0 40 0 40,scale=0.45]{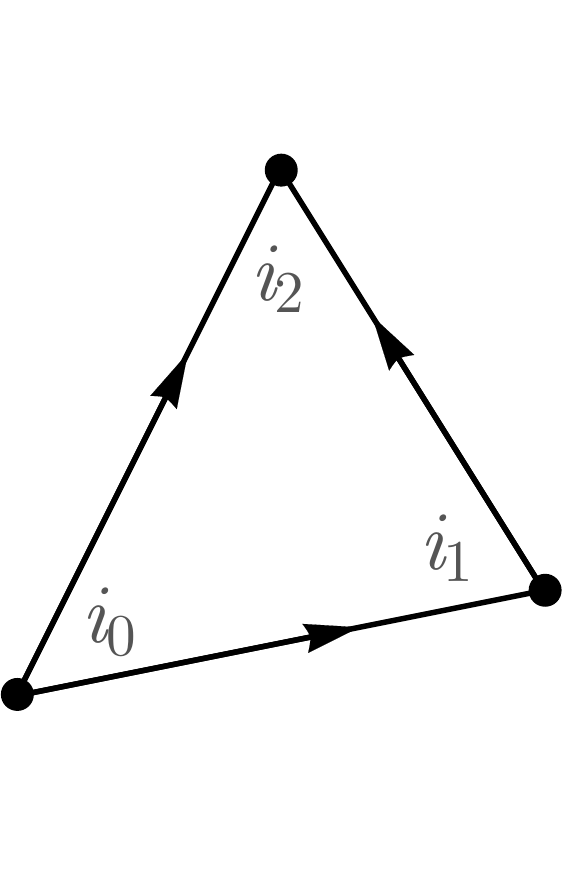}\hspace{5mm}
\includegraphics[clip=true, trim = 0 40 0 40,scale=0.45]{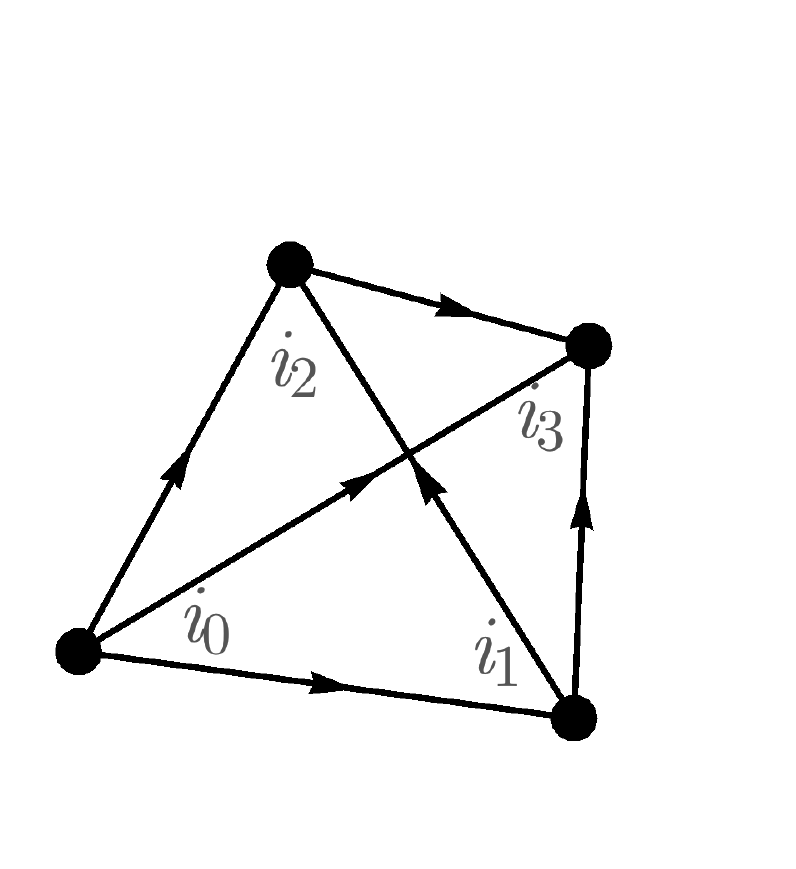}
\end{center}
\caption{Ordering of vertices within a branched 2-simplex and a branched 3-simplex. \label{fig:tritetord}}
\end{figure}

We now use the elements introduced above to write down a general Floquet SPT Hamiltonian. For a $d$-dimensional simplicial complex with a branching structure, we label the vertices within each $d$-simplex as $\{i_0,i_1,\ldots,i_{d}\}$, ordered according to how many arrows (associated with the branching structure) point towards them. Fig.~\ref{fig:tritetord} shows the explicit vertex ordering for a 2-simplex and a 3-simplex. 

As before, we associate the state on each vertex with an element $g$, which belongs to the regular representation of group $G$. We write the states on a $d$-simplex as $\ket{g_{i_0},g_{i_1},\ldots,g_{i_d}}$, where $g_{i_0}$ gives the state at vertex $i_0$, etc. With each such $d$-simplex, we associate the following Hermitian operator,
\beq
\h_{i_0,i_1,\ldots,i_d}&=&\sum_{g_{i_0},\ldots,g_{i_d}}\left(-1\right)^{s_{\{i_0,\ldots,i_d\}}}f\left(g_{i_0},\ldots,g_{i_d}\right)\nonumber\\
&&\ket{g_{i_0},g_{i_1},\ldots,g_{i_d}}\bra{g_{i_0},g_{i_1},\ldots,g_{i_d}},
\eeq
where $f\left(g_{i_0},\ldots,g_{i_d}\right)$ is the unique number in the range $\left(-\pi,\pi\right]$ that satisfies
\beq
e^{if\left(g_{i_0},\ldots,g_{i_d}\right)}&=&\nu_d\left(g_{i_0},\ldots,g_{i_d}\right),
\eeq
and $s_{\{i_0,\ldots,i_d\}}$ is an orientation factor that is zero if the vertices $\{i_0,\ldots,i_d\}$ have the same orientation as the manifold $M$, and is one if the vertices have the opposite orientation to $M$.

The operator $\h_{i_0,i_1,\ldots,i_d}$ has a number of useful properties. First, it commutes with the global actions of the group $G$. To see this, we write the global action of the group element $g$ as $V(g)$ and find
\begin{widetext}
\beq
V(g)\h_{i_0,i_1,\ldots,i_d}V(g)^{-1}&=&\sum_{g_{i_0},\ldots,g_{i_d}}\left(-1\right)^{s_{\{i_0,\ldots,i_d\}}}f\left(g_{i_0},\ldots,g_{i_d}\right)\ket{gg_{i_0},\ldots,gg_{i_d}}\bra{gg_{i_0},\ldots,gg_{i_d}}\\
&=&\sum_{g_{i_0},\ldots,g_{i_d}}\left(-1\right)^{s_{\{i_0,\ldots,i_d\}}}f\left(gg_{i_0},\ldots,gg_{i_d}\right)\ket{gg_{i_0},\ldots,gg_{i_d}}\bra{gg_{i_0},\ldots,gg_{i_d}},\nonumber
\eeq
where in the second line we have used the relation in Eq.~\eqref{eq:cochainrel}. Since the sum is over all group elements, the final expression is equivalent to $\h_{i_0,i_1,\ldots,i_d}$.

A second useful property of $\h_{i_0,i_1,\ldots,i_d}$ is that it may be rewritten in the form 
\beq
\h_{i_0,i_1,\ldots,i_d}=\left(\sum_k{\h_k}\right)\mod{2\pi},
\eeq
with
\beq
\h_k&=&\sum_{g_{i_0},\ldots,g_{i_d}}\left(-1\right)^{s_{\{i_0,\ldots,i_d\}}}\left(-1\right)^kh_k\left(g_{i_0},\ldots,\hat{g}_{i_k},\ldots,g_{i_d}\right)\ket{g_{i_0},g_{i_1},\ldots,g_{i_d}}\bra{g_{i_0},g_{i_1},\ldots,g_{i_d}}.
\eeq
\end{widetext}
In the above, the caret indicates that $g_{i_k}$ is omitted from the arguments of $h_k$, and $h_k$ is the unique number in the range $(-\pi,\pi]$ that satisfies
\bequ
e^{ih_k\left(g_{i_0},\ldots,\hat{g}_{i_k},\ldots,g_{i_d}\right)}=\nu_d(1_G,g_{i_0},\ldots,\hat{g}_{i_k},\ldots,g_{i_d}),
\eequ
where $1_G$ is the identity element in $G$. The equivalence between these two expressions for $\h_{i_0,i_1,\ldots,i_d}$ uses a well-known result from group cohomology theory, and is proved in Ref.~\cite{Chen:2013foa}. We note that $\h_k$ is independent of the state $g_{i_k}$ on vertex $i_k$, and so we may formally regard it as acting on the $d-1$ simplex $\{i_0,\ldots,\hat{i}_k,\ldots,i_{d}\}$.

For the Hamiltonian of the complete system, we sum over all $d$-simplices (indexed by $p$) to obtain
\bequ
\h_{G,d}=\sum_{p}\h^p_{i_0,i_1,\ldots,i_d}\equiv\sum_{p}\left[\left(\sum_k{\h^p_k}\right)\mbox{mod~}{2\pi}\right].\label{eq:cohomology_hamiltonian_gend}
\eequ

\subsection{Nontrivial Unitary Evolution}
\begin{figure}[h!]
\begin{center}
\includegraphics[clip=true, trim = 40 30 20 60,scale=0.45]{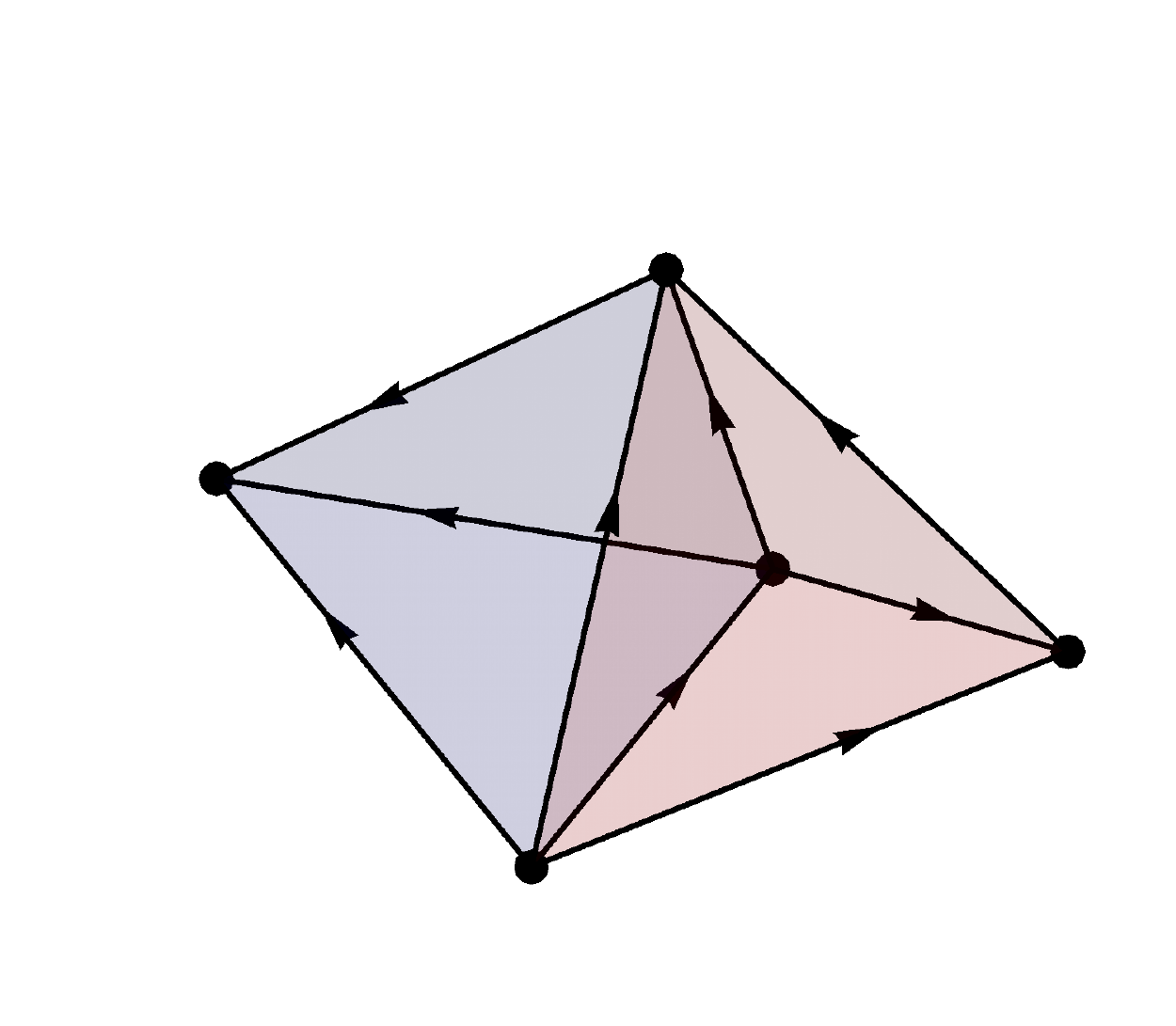}
\includegraphics[clip=true, trim = 40 30 80 60,scale=0.45]{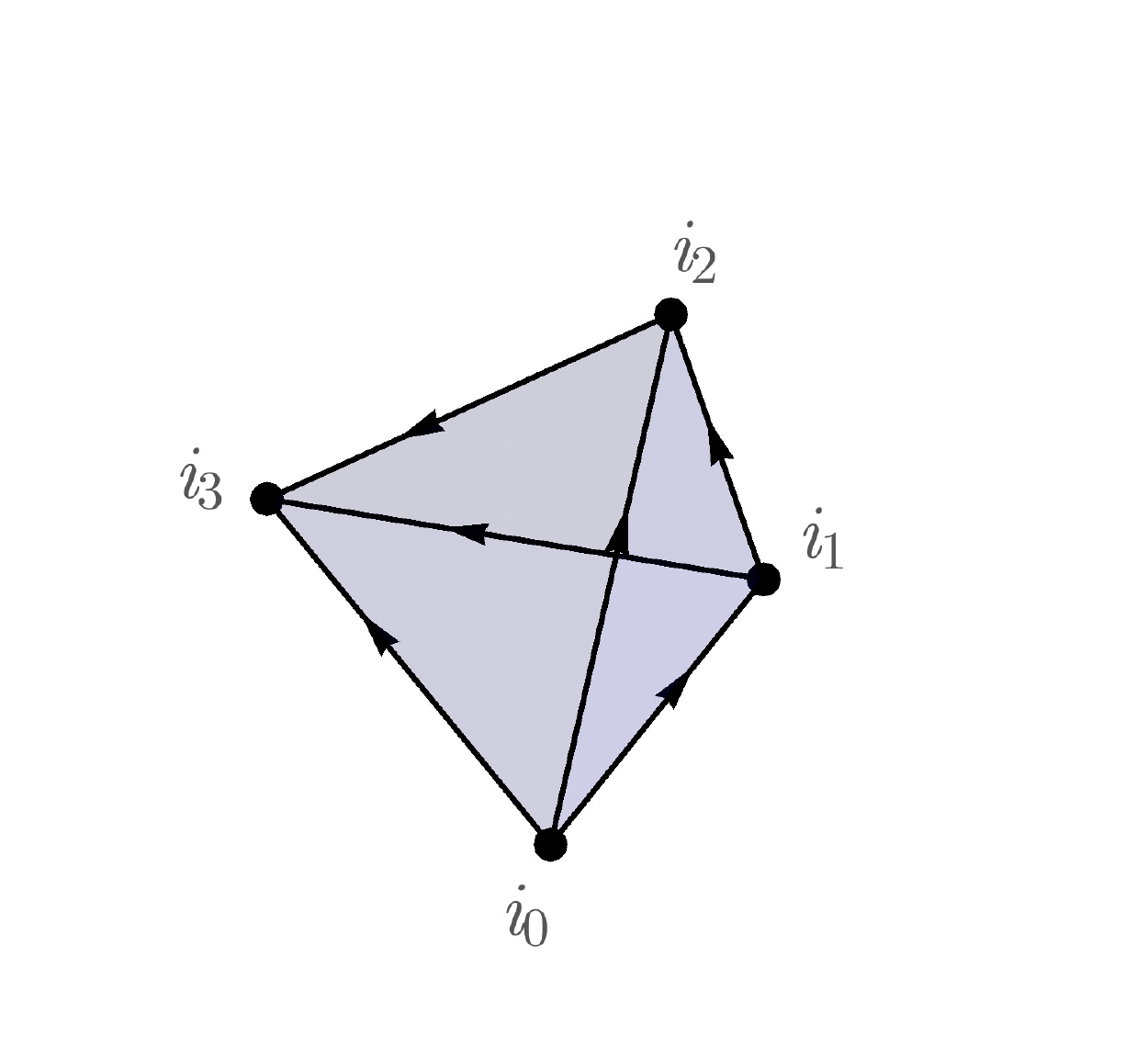}
\includegraphics[clip=true, trim = 130 30 20 60,scale=0.45]{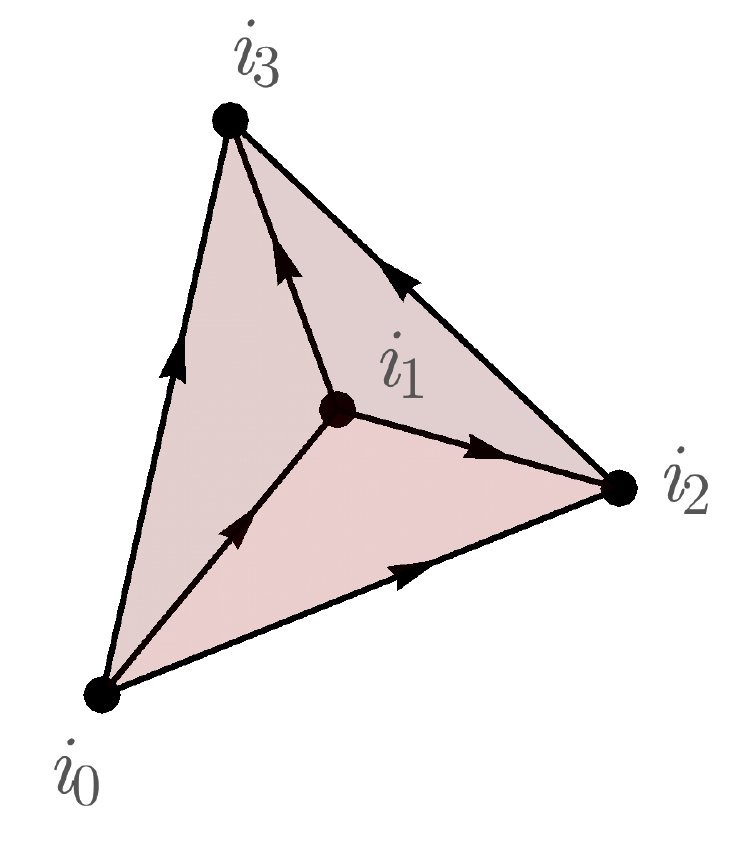}
\end{center}
\caption{Two 3-simplices that share a common 2-simplex. Since both 3-simplices have the same orientation (which we take to be positive), the contributions to $U_{G,3}(1)$ from the shared 2-simplex will be $\nu^{-1}_2(1_G,g_{i_0},g_{i_1},g_{i_2})$ from the left 3-simplex and $\nu_2(1_G,g_{i_0},g_{i_1},g_{i_3})$ from the right 3-simplex, which cancel.\label{fig:simplex_cancellation}}
\end{figure}

We now show that this Hamiltonian generates a nontrivial unitary evolution when evolved until $t=1$ to give $U_{G,d}(1)=e^{-i\h_{G,d}}$. First, it is easy to verify that for every $(d-1)$-simplex that is shared by two $d$-simplices, the two terms in $U_{G,d}(1)$ corresponding to this $(d-1)$-simplex cancel out. This is shown pictorially in Fig.~\ref{fig:simplex_cancellation}. Thus, for a closed system, $U_{G,d}(1)=\id$.

For an open system, however, there will be a $(d-1)$-dimensional boundary consisting of $(d-1)$-simplices whose terms in $U_{G,d}(1)$ do not cancel. Acting on a product state $\ket{\{g\}}$, the open system unitary operator will have the action
\beq
U_{G,d}(1)\ket{\{g\}}&=&\prod_{\{i_0,\ldots,i_{d-1}\}}\nu_{d}\left(g_{i_0},\ldots,g_{i_{d-1}}\right)\ket{\{g\}}\nonumber\\
&\equiv& \left( \id_{\rm bulk} \otimes U_{\rm eff} \right)\ket{\{g\}},
\eeq
where the product is only over the $(d-1)$-simplices at the boundary. 

We now specifically consider a trivial product state, which may be written as $\ket{\psi} = \ket{\psi_{\rm bulk}} \otimes \ket{\psi_{d-1}}$, where $\ket{\psi_{d-1}}$ is the trivial symmetric product state on the $(d-1)$-dimensional boundary of size $N$, 
\beq
\ket{\psi_{d-1}}&=&\sum_{g_1,\ldots,g_{N}}\ket{g_1,\ldots,g_{N}}.
\eeq
The action of the unitary evolution produces the state $U_{G,d}(1)\ket{\psi}=\ket{\psi_{\rm bulk}} \times \ket{\psi_{\nu_{d-1}}}$, where $\ket{\psi_{\nu_{d-1}}}$  is an SPT state associated with $\nu_{d-1}$. The effective edge unitary $U_{\rm eff}$ is thus an operator that transforms a trivial SPT state into a nontrivial one: it therefore cannot be obtained through a local unitary evolution with a Hamiltonian that preserves the symmetry of $G$ (but it can be generated by a Hamiltonian that is not symmetric). Each such $U_{\rm eff}$ associated with a distinct cohomology class produces a distinct local unitary evolution of this form, and acts on a $(d-1)$-dimensional SPT state at the edge to produce an SPT state in a different topological phase. 

While the group cohomology classification of static SPT states remains a conjecture, and there are indeed states that lie outside of this classification \cite{Burnell:2014jm}, there are a number of arguments which show that the states corresponding to each cohomology class are distinct and describe different SPT phases. In addition to the construction presented in Ref.~\cite{Chen:2013foa}, Ref.~\cite{Chen:2012je} showed that the same classification could be obtained by studying the anomalous action of the symmetries on the edge. Further, in Refs.~\cite{Wang:2016gj,Wang:2015if,Levin:2012jq}, the authors showed that gauging distinct group cohomology-based SPT phases gives rise to phases with different forms of topological order. On this basis, they argued that the corresponding SPT states could not be deformed to one another, further supporting the idea that, even if incomplete, the group cohomology classification produces a set of distinct phases. 

Our group cohomology construction gives rise to effective edge unitaries that relate trivial SPT phases to nontrivial ones. Thus, any rigorous argument showing that a particular static SPT phase is distinct from another implies that the corresponding effective edge unitaries are also inequivalent (in the sense introduced in Sec.~\ref{sec:nontrivial_unitary_loops}). The arguments of Appendix~\ref{app:loop_equivalence} then show further that the corresponding bulk drives for distinct elements of $H^d[G,{\rm U(1)}]$ are also inequivalent and cannot be continuously deformed to one another. Thus, while the group cohomology classification proposed in this paper need not be complete, the corresponding classification of static SPT phases allows us to argue that each distinct cohomology element defines a distinct FSPT phase, with the same level of confidence as for the static case.

We note that our work also provides a new perspective on static SPT phases: The SPT states classified by group cohomology may be obtained from trivial SPT phases through a local unitary evolution that commutes with the symmetry generators, but it is known that this unitary evolution cannot be generated by a local symmetric Hamiltonian in any finite time (i.e. it cannot be generated by a symmetric local quantum circuit) \cite{Huang:2015gt}. In our work, through explicit construction, we have shown that these states may be generated from a product state by a symmetric finite-depth circuit at the boundary of a higher-dimensional system. This may be one route to generating SPT states experimentally.
\subsection{Summary of Nontrivial Loop Order\label{sec:loop_order}}
In the preceding sections we introduced a general set of FSPT drives applicable to any symmetry group $G$ on any manifold and in any number of dimensions. Specifically, there exits a unitary evolution $U^{\nu_d}_{G,d}(t)$, corresponding to each element from the cohomology group $H^d[G,{\rm U(1)}]$, which generates a nontrivial loop. Since elements from $H^d[G,{\rm U(1)}]$ form an Abelian group, the unitary evolutions $U^{\nu_d}_{G,d}(t)$ also form an Abelian group---we can run several drives in sequence, and the resulting FSPT drive can be derived from the structure of $H^d[G,{\rm U(1)}]$. Any FSPT loop in the same class as   one of the group cohomology drives constructed above can be obtained from it through a symmetric evolution at the edge, and may therefore be labeled by the relevant cohomology element.

More generally, a nontrivial loop may also have order (symmetry-protected or topological) that lies outside of the group cohomology construction, as discussed in Sec.~\ref{sec:TSDs} and Sec.~\ref{sec:FSPT_not_cohomology}. The action of the drives constructed in Sec.~\ref{sec:FSPT_not_cohomology} differ from the group cohomology drives in that they seem to map trivial product states onto other trivial product states. We leave a detailed investigation of how these different types of order combine to future work.
\section{Endpoint Unitaries and Dynamical Phases\label{sec:W_standardpaths}}

Having introduced a series of nontrivial loops, we now discuss these evolutions in the context of homotopy, using the ideas introduced in Sec.~\ref{sec:phase_structure_u}. 

We recall that in order to obtain a meaningful classification of unitary evolutions within the space $S_0$ or $S_G$ using homotopy, we must restrict the set of endpoints to a region $W$. To be a useful restriction, there should be no nontrivial unitary loops contained wholly within $W$.

The simplest restriction takes $W$ as a single point, $P$. In this case, the relative order of any pair of unitary evolutions, $U_1$ and $U_2$, that end at $P$ can be obtained by calculating the loop order of the composition $U_2^{-1}\circ U_1$. Within the space $S_G$, the nontrivial loops will have order as described in Sec.~\ref{sec:loop_order}.

A more useful choice for $W$ would allow us to compare unitary evolutions with different endpoints. Consider first the space $S_0$. From the explicit examples that we know of loops in this space, it seems that such loops all include points during the evolution where the unitary operator cannot be expressed as $e^{-iH}$ for any local Hamiltonian $H$. This leads us to conjecture that the space $W$ of endpoint unitaries which \emph{can} be expressed as $U(1)=e^{-iH_F}$ for a local Floquet Hamiltonian $H_F$ does not have any nontrivial loops. From our previous discussion, it then follows that there is a well-defined notion of relative order between unitary evolutions that end in the region $W$. If we also define the standard drive to the endpoint $U(1)=e^{-iH_F}$ as that which is generated by the constant Hamiltonian $H_F$, then there is also a notion of absolute order for these unitary evolutions. The phases thus defined are stable to perturbations which keep the endpoints of the unitary evolutions within this space.   

These considerations also immediately apply to the subspace of unitary evolutions in $S_0$ that have endpoints of the form $U(1)=e^{-iH_{\rm MBL}}$, where $H_{\rm MBL}$ is an MBL Hamiltonian. These are of experimental interest since MBL may be used to avoid the problem of heating in Floquet systems \cite{DAlessio:2013fv,Lazarides:2014ie,Abanin:2014te,Ponte:2015hm,Ponte:2015dc,Lazarides:2015jd,Abanin:2015bc,Khemani:2016gd,Zhang:2016vt,Zhang:2016tb} and may also aid in the detection of edge modes \cite{Chandran:2014dk,Bahri:2015ib}. However, it has not been proven that MBL is stable to heating in dimensions $d>1$ \cite{DeRoeck:2016us,Agarwal:2016vk}, and for this reason, it is perhaps useful to separate out those topological aspects which can be framed independently of assumptions about MBL. Nevertheless, a generic MBL unitary is expected to remain in the MBL phase when perturbed by arbitrary local perturbations, and so MBL unitaries lend themselves to a particularly robust notion of phase.

Another useful choice for $W$ within $S_0$ is the set of unitaries that correspond to time crystals, which are not MBL but which have an \emph{almost} complete set of local integrals of motion. Again, based on the form of the unitary evolution at intermediate points, we conjecture that this choice of $W$ is also free of nontrivial loops. Further, the `time-crystal order' of unitary evolutions in this space has  been argued to be stable to arbitrary local perturbations~\cite{vonKeyserlingk:2016ev,Else:2016gf}. Note that these considerations also apply to the chiral anomalous drives discussed in Refs.~\cite{Po:2016uwb,Harper:2016us}, and not just the symmetric ones discussed here.

We now consider the symmetry-protected space $S_G$. We first note that if $W$ is the set of all unitary evolutions whose endpoints have MBL Hamiltonians (with symmetry $G$) then this space certainly has nontrivial group cohomology loops. This, in particular, means that the space of unitary evolutions with endpoints of the form $U(1)=e^{-iH_F}$, for some local Floquet Hamiltonian $H_F$, also has nontrivial loops. To illustrate this point, consider a version of the SPT Hamiltonian in Eq.~\eqref{eq:cohomology_hamiltonian_gend}, where each term now comes with a random coefficient,
\beq
\h_{G,d}^{{\rm MBL}}=\sum_{p}A^p\h^p_{i_0,i_1,\ldots,i_d}.
\eeq
The set $\{A^p\}$ may be taken, for example, from a standard Normal distribution, so that $\h_{G,d}^{{\rm MBL}}$ is an MBL Hamiltonian (subject to the general concerns about the stability of MBL in $d>1$ \cite{DeRoeck:2016us,Agarwal:2016vk}). 

Now, since all eigenstates of this Hamiltonian may be written in the form $\ket{g_1,\ldots,g_N}$, they must come in degenerate multiplets related by the action of the global group symmetry operators. Thus, they exhibit a version of spontaneous symmetry breaking, as may be explicitly verified in the 2d $\zbb_2\times\zbb_2$ case (and as was noted for the 1d case in Ref.~\cite{vonKeyserlingk:2016ea}). By considering the path of MBL Hamiltonians $\h_{g,d}^{\rm MBL}(s)$, where changing $s$ alters the coefficients $A^p$ according to $A^p(s)=A^p(0)+s$, we see that there are nontrivial loops in this space of MBL Hamiltonian endpoints. Explicitly, taking $s$ from zero to one yields the unitary evolution $U=e^{-i\h_{G,d}^{{\rm MBL}}}U_{G,d}(1)$, where $U_{G,d}(1)$ is the nontrivial loop given in Sec.~\ref{sec:cohomology_anyd}. 

Note, however, that for the groups of type ${\rm U(1)\times U(1)}$, for which we constructed FSPT drives outside the group cohomology paradigm, we don't expect any nontrivial loops in this space. This is based on the discussion of the space $S_0$ given above. Thus, this space of endpoints (and so also the space of endpoint MBL Hamiltonians with the group symmetry) may each be used to define dynamical phases of this type.

From the structure of the nontrivial loops, as well as the arguments of \cite{Else:2016ja}, one is led to believe that there are no nontrivial group cohomology loops in the space of MBL Hamiltonians without spontaneous symmetry breaking (with or without non-trivial SPT order). We can therefore take $W$ to be the set of endpoint unitaries given by $U=e^{-iH_{\rm SPT, MBL}}$, where $H_{\rm SPT, MBL}$ is an SPT-ordered MBL Hamiltonian without spontaneous symmetry breaking. Each class of SPT-ordered MBL Hamiltonians defines a distinct and disconnected space of endpoints $W$. Similar considerations also apply to a set of unitary evolutions whose endpoints have the form $e^{-iH}V(T)$, where $V(T)$ is a translation operator and $e^{-iH}$ commutes with $V(T)$, as arises in the proposed drives of Ref.~\cite{Else:2016ja}. If the space defined by endpoints $U(1)=e^{-iH}$ has no loops, then the space of endpoints of the form $U(1)=e^{-iH}V(T)$ does not either.

Finally, we could also choose $W$ to be the space of endpoint MBL Hamiltonians in $S_G$ with some particular type of topological order. Indeed, distinct classes of MBL Hamiltonians of this form, such as those corresponding to symmetry enriched topological phases, give rise to distinct disconnected spaces of endpoints with no loops. For each of these spaces, the existence of well-defined Floquet Hamiltonians permits the definition of absolute order in addition to the definition of relative order.

This is far from an exhaustive list of possible choices of $W$. The utility of our approach is that any choice of $W$ may be handled within this framework.

\section{Conclusion\label{sec:conclusion}}
In this work, we have discussed bosonic Floquet topological phases with symmetry in all dimensions. We began with some general homotopic considerations about the topology of the phase space of unitary evolutions and argued that, in general, two types of Floquet order may be identified. The first is related to the bulk eigenstate order of the Floquet unitary at the end of the evolution, while the second is related to the form of the complete bulk evolution (and not just the end point). The latter is an intrinsic dynamical order which has no static analog. 

To define phases, we considered families of unitary evolutions with end points in some fixed space, $W$ and argued that these were stable to various classes of perturbation depending on the nature of the space $W$. Within this framework, two types of order could be discussed: a relative order and an absolute order, each based on the notion of loops (drives whose Floquet Hamiltonian is zero). 

We constructed a set of nontrivial loops based on group cohomology in all dimensions, which associates a type of dynamical FSPT order with each element of the cohomology group $H^d[G,{\rm U(1)}]$. These loops were shown to have nontrivial effective unitaries at the boundary, which map trivial product states at the edge onto nontrivial SPT states. An effective edge unitary of this kind cannot be obtained from a local, finite-time, symmetric evolution (i.e. from a local symmetric quantum circuit). We argued further that a loop cannot be transformed continuously (in the space of symmetric loops) to another unless their effective edge unitaries are related through a local symmetric unitary evolution.

We also discussed a set of topological symmetric drives based on the anomalous chiral drives of Ref.~\cite{Po:2016uwb,Harper:2016us}. These exhibit chiral information transport at the boundary of the open system, and cannot be deformed to the trivial loop even in the absence of symmetry restrictions. This motivated the construction of a class of FSPT drives that lie outside of the group cohomology paradigm.

Although we defer a detailed comparison to Appendix~\ref{sec:eigenstate_order}, we note here that our work has a number of differences from other approaches in the literature (based on bulk eigenstate order). Notably, within our framework, bulk eigenstate order does not seem to be associated with (and cannot uniquely identify) the dynamical topological loop order, in sharp contrast to static systems. However, the notion of pumping in two dimensions proposed in Refs.~\cite{vonKeyserlingk:2016bq,Potter:2016tba,Else:2016ja,Potter:2016vq} has a direct analog in our work generalized to all dimensions, related to the generation of nontrivial SPT states at the boundary. It would be interesting to further compare these approaches.

Our work also provides an interesting perspective on static SPT phases. While SPT states cannot be generated from trivial product states by finite-depth symmetric local quantum circuits \cite{Huang:2015gt}, here we have shown (through explicit construction) that these states may be generated from a product state by a symmetric local unitary evolution at the boundary of a higher dimensional system. This may be one route to generating these SPT states experimentally.

The stability of MBL systems to local perturbations is likely to transfer to unitary evolutions, and so one expects there to be analogs of static MBL phases (SPT, spontaneous symmetry breaking phases, etc.) in driven systems. In addition to these, new phases of eigenstate order not present in static systems may also arise, such as spontaneous breaking of discrete time translation symmetry in Floquet time crystals. For some of these classes of stable unitary evolutions, our construction of nontrivial loops shows that there exists a stable form of inherently dynamical order, manifested as a relative or absolute order (or both) depending on the space under consideration. The nature of the transitions between novel phases of this kind is an interesting subject for future study. 

Localization issues are the most pertinent for any experimental realization of these novel phases. While we have made various conjectures about the existence of nontrivial loops in a set of different spaces of MBL unitary endpoints, it would be useful to substantiate these with numerical studies or further theoretical investigation. In principle, the action of an effective edge unitary may be detected by studying the switching of the eigenstate order of the corresponding edge modes. However, this involves the challenge of detecting closed SPT systems, for which the proposals of Ref.~\cite{Bahri:2015ib}, for instance, do not apply.

\begin{acknowledgements}
We are grateful to A.~Kitaev, A.~C.~Potter, L.~Fidkowski, X.~Chen and S.~L.~Sondhi for useful discussions on a range of topics related to the current manuscript. R.~R. and F.~H. acknowledge support from the NSF under CAREER DMR-1455368 and the Alfred P. Sloan foundation.
\end{acknowledgements}

\appendix

\section{Independence of Effective Edge Unitary on Choice of Boundary\label{app:boundary_choice}}
In this appendix, we show that the effective edge unitary of a nontrivial drive is independent of the choice of boundary for the open system (in terms of its topological classification).  As usual, we assume that the unitary evolution of the system is generated by some symmetry-respecting Hamiltonian $H(t)$, which is a sum of local terms. For a nontrivial loop, the unitary operator of a closed system is the identity. To obtain the unitary evolution for the open system, we simply exclude terms from $H(t)$ that connect sites either side of the boundary cut.

\begin{figure}[t]
\begin{center}
\includegraphics[scale=0.35]{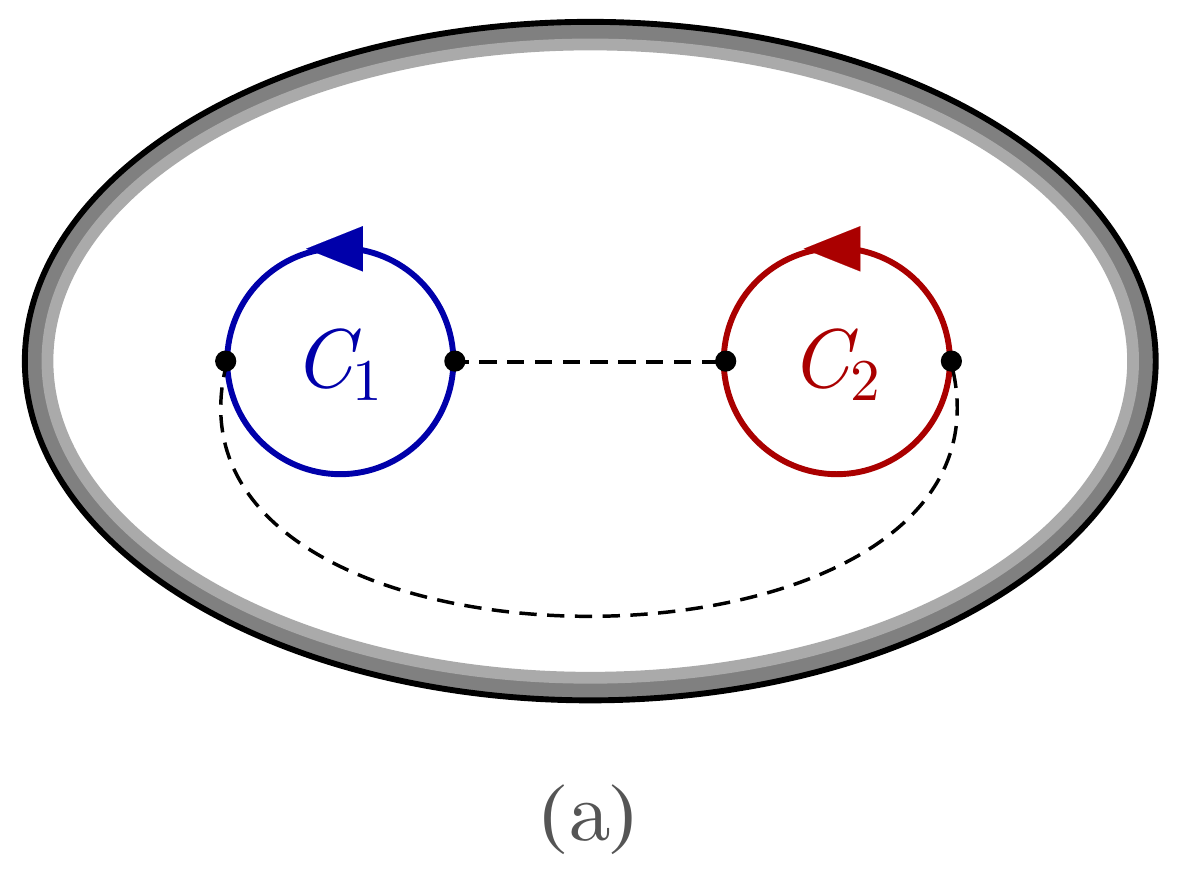}\hspace{5mm}
\includegraphics[clip = true, trim = 10 80 0 0,scale=0.32]{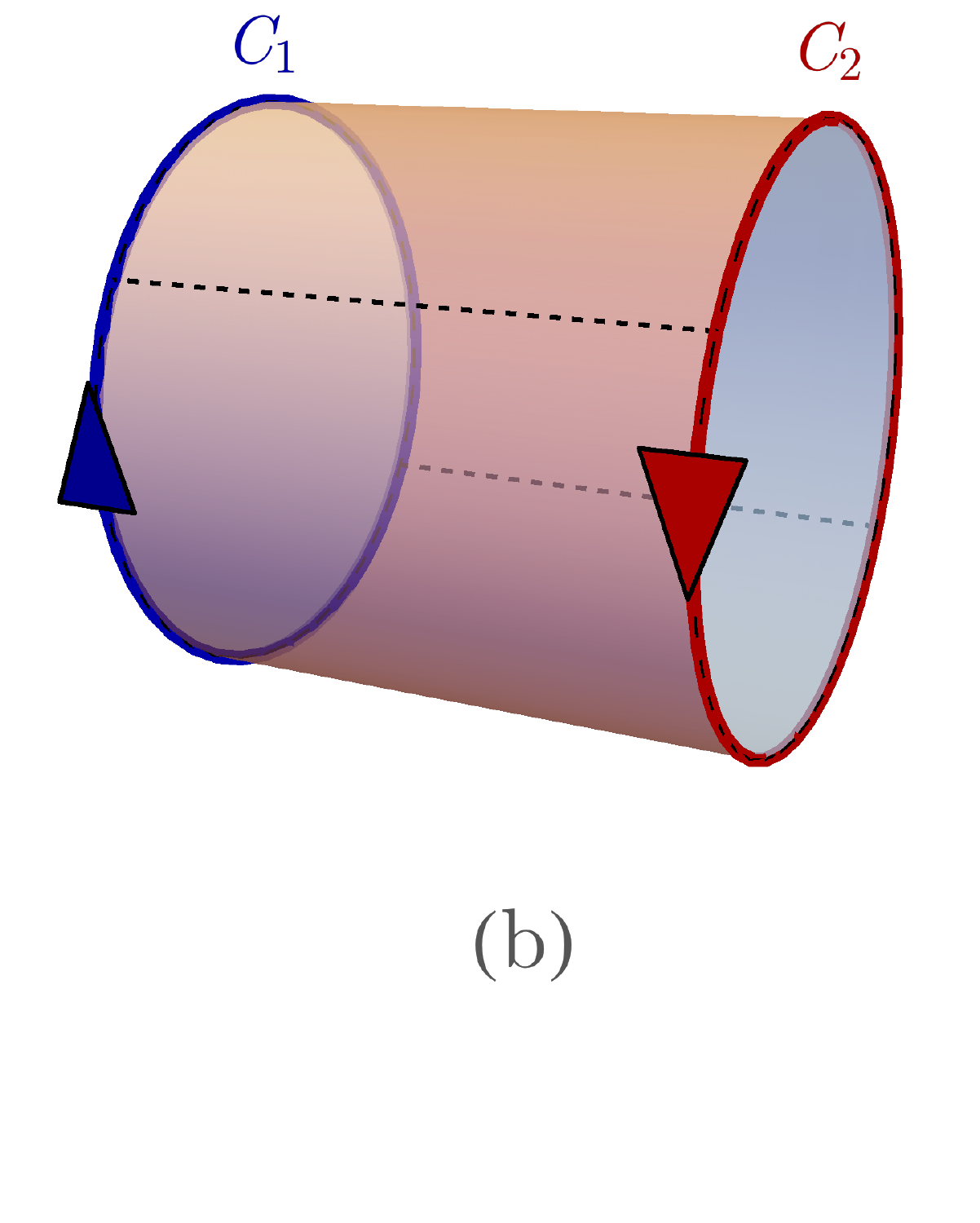}
\end{center}
\caption{(a) and (b): Two boundaries $C_1$ and $C_2$ of an otherwise closed system, (a) on a generic 2d surface and (b) after being deformed into a cylinder. The edge Hamiltonian $H_{\rm edge}(t)$ connects sites from each boundary, indicated by dashed lines. \label{fig:boundaries}}
\end{figure}

We consider a specific open system with two oriented boundaries $C_1$ and $C_2$, which we assume are of equal size (see Fig.~\ref{fig:boundaries}). We may reconnect these two boundaries by adding terms to $H(t)$ that connect sites from each oriented boundary. This could be interpreted as deforming the system into a cylinder, and then connecting the two ends to form a torus. With this interpretation, $C_1$ and $C_2$ have orientations as shown in Fig.~\ref{fig:boundaries}. 

Specifically, we write the new Hamiltonian as
\beq
H_{\rm closed}(t)&=&H_{\rm open}(t)+H_{\rm edge}(t),
\eeq
in terms of the Hamiltonian of the previous open system and the Hamiltonian $H_{\rm edge}(t)$, which contains terms that connect $C_1$ to $C_2$. The evolution of this new, closed, system must be the identity, since we have assumed that the unitary evolution generated by $H(t)$ is a nontrivial loop.

Borrowing arguments from Ref.~\cite{Harper:2016us}, we may relate the evolutions of the closed and open systems through
\beq
U_{\rm open}(t)&=&U'_{\rm edge}(t)U_{\rm closed}(t)\equiv U'_{\rm edge}(t),
\eeq
where $U_{\rm open/closed}(t)$ is generated by $H_{\rm open/closed}(t)$ and $U'_{\rm edge}(t)$ is generated by $H'_{\rm edge}(t)$, a Hamiltonian localized at the boundary between $C_1$ and $C_2$ that is obtained from $H_{\rm edge}(t)$ through conjugation with $U_{\rm closed}$ (see Ref.~\cite{Harper:2016us}).

Since every term in the Hamiltonian respects the symmetry $G$, and since $H_{\rm edge}(t)$ is a quasi-1d Hamiltonian, the unitary $U'_{\rm edge}(t)$ cannot generate nontrivial 1d SPT order, and so $U'_{\rm edge}$ is trivial. Furthermore, since $H_{\rm open}(t)$ only differs from $H_{\rm closed}(t)$ away from $C_1$ and $C_2$, $U'_{\rm edge}(t)$ must be localized near these regions (from the Lieb-Robinson bound on the propagation of information). In this way, we may write
\beq
U'_{\rm edge}(1)&=&U_{C_1}\otimes U_{C_2},
\eeq
where $U_{C_j}$ only has support in the neighborhood of the region $C_j$.

Since $U'_{\rm edge}(1)$ is trivial, its anomalous action near the left edge (in the cylindrical interpretation) must be equal to the anomalous action near the right edge, but with the opposite orientation. However, it is evident from Fig.~\ref{fig:boundaries}(b) that after deforming the surface into a cylinder, the boundaries $C_1$ and $C_2$ do have opposite orientation. In this way, we find that any choice of $C_1$ or $C_2$ yields the same anomalous action. 

For a large enough system, and for large enought boundaries $C_1$ and $C_2$, small deformations in the size of the boundary may be accommodated through local changes in the Hamiltonian at the edge, which cannot affect the above conclusions. In this way, the argument is also valid for $C_1$ and $C_2$ of different size, provided they are large enough. While the argument presented above is for two-dimensional systems, a $d$-dimensional generalization shows that the same conclusion holds in higher dimensions as well.

\section{Homotopy of Loops and Equivalence of Effective Edge Unitaries \label{app:loop_equivalence}}
In this appendix, we show that if two loops, $L_1$ and $L_2$, are homotopic to each other, then their effective edge unitaries, $U_{\rm eff}^1$ and $U_{\rm eff}^2$ are equivalent up to local symmetric perturbations. By equivalence, we say that $U_{\rm eff}^1$ and $U_{\rm eff}^2$ are equivalent if and only if
\beq
U_{\rm eff}^1= VU_{\rm eff}^2,\label{eq:ueff_equivalence}
\eeq
where $V$ is a symmetric unitary evolution localized at the edge. Homotopy is defined as in Sec.~\ref{sec:phase_structure_u}. Specifically, $L_1$ and $L_2$ are homotopic if and only if there exists a (symmetric, local) homotopy Hamiltonian, $H(s,t)$, where $H(0,t)$ generates $L_1$ and $H(1,t)$ generates $L_2$, such that the end point of the evolution remains within some fixed region of endpoints $W$ for all $0\leq s\leq1$. (For example, $W$ may be the set of endpoints of the form $U=e^{-iH_{\rm MBL}}$, where $H_{\rm MBL}$ is an MBL Hamiltonian. See Sec.~\ref{sec:phase_structure_u} for a full discussion of the phase space of unitary evolutions).

With this definition, the homotopy Hamiltonian $H(s,t)$ defines a homotopy unitary operator $U(s,t)$, which we may expand at $t=1$ to obtain
\bequ
U(s+\epsilon,1)=\lim_{\Delta t\to 0}\left[e^{-i\tilde{H}(1)\Delta t}e^{-i\tilde{H}(1-\Delta t)\Delta t}\ldots e^{-i\tilde{H}(\Delta t)\Delta t}\right],
\eequ
where $\tilde{H}(t)$ is a shorthand for $H(s+\epsilon, t)$. We now write
\beq
H(s+\epsilon,t)&=&H(s,t)+H'(s,t),
\eeq
where $H'$ can be made as small as desired by making $\epsilon$ arbitrarily small. We may then write the homotopy unitary as
\beq
U(s+\epsilon,1)&=&\lim_{\Delta t\to 0}\left[e^{-iH(s,1)\Delta t}e^{-iH'(s,1)\Delta t}\right.\nonumber\\
&&\left.\ldots e^{-iH(s,\Delta t)\Delta t}e^{-iH'(s,\Delta t)\Delta t}\right]\\
&=&U'(\epsilon,1)U(s,1),\nonumber
\eeq
where $U'(\epsilon,1)$ is some local unitary (not necessarily that which is generated by $H'(s,t)$), and in the second line we have used a similar rearrangement to that given in Ref.~\cite{Harper:2016us}.

Now, since $U(s,1)$ and $U(s+\epsilon,1)$ are both symmetric, it follows that
\beq
U'(\epsilon,1)= U(s+\epsilon,1)U(s,1)^{-1}
\eeq
is also symmetric and, further, that it is connected to the identity as $\epsilon\to0$. Since this holds for every infinitesimal step of the homotopy, it follows that $U(0,1)=U_{\rm eff}^1$ and $U(1,1)=U_{\rm eff}^2$ are equivalent in the sense of Eq.~\eqref{eq:ueff_equivalence}.


\section{Projective Representations of $\zbb_2\times\zbb_2$\label{app:z2z2}}
The group $\zbb_2\times\zbb_2$ is an Abelian group of order four with presentation
\beq
\langle a,b|a^2=b^2=(ab)^2=1\rangle.
\eeq
It is the direct product of two copies of the cyclic group $\zbb_2$, and in a physical setting, may be associated with a system of two Ising spins. The group multiplication table is shown in Tab.~\ref{tab:z2z2details}(a).

In addition to the linear representations, the group $\zbb_2\times\zbb_2$ also has a nontrivial projective representation, which may be represented by the Pauli matrices \cite{Zeng:2015vc}
\beq
P_\omega(1)&=&\id\\
P_\omega(a)&=&\sigma^x\nonumber\\
P_\omega(b)&=&\sigma^z\nonumber\\
P_\omega(ab)&=&\sigma^y.\nonumber
\eeq
It may be verified that this is a projective representation of the group, as defined in Sec.~\ref{sec:projrep}, and that the corresponding factor system takes the values given in Tab.~\ref{tab:z2z2details}(b). The related U(1) $\alpha$ phases defined in Eq.~\ref{eq:alphadef} are also given in Tab.~\ref{tab:z2z2details}(c).

\begin{table*}[t]
\be
\begin{array}{ccccc}
\begin{array}{c|cccc}
* & 1 & a & b & ab \\
\hline
1 & 1 & a & b & ab\\
a & a & 1 & ab & b\\
b & b & ab & 1 & a\\
ab & ab & b & a & 1.
\end{array}
&&
\begin{array}{c|cccc}
\omega(g_a,g_b) & g_b= 1 & a & b & ab \\
\hline
g_a=1 & 1 & 1 & 1 & 1 \\
a & 1 & 1 & -i & i \\
b & 1 & i & 1 & -i \\
ab & 1 & -i & i & 1
\end{array}
&&
\begin{array}{c|cccc}
\alpha(g_a,g_b) & g_b= 1 & a & b & ab \\
\hline
g_a=1 & 1 & 1 & 1 & 1 \\
a & 1 & 1 & i & -i \\
b & 1 & -i & 1 & i \\
ab & 1 & i & -i & 1
\end{array}\\
(a) && (b) && (c)
\end{array}
\ee
\caption{(a) Group multiplication table for $\zbb_2\times\zbb_2$. (b) factor system and (c) U(1) $\alpha$ phases for the nontrivial projective representation of $\zbb_2\times\zbb_2$.\label{tab:z2z2details}}
\end{table*}


\section{Eigenstate Order and Floquet Phases\label{sec:eigenstate_order}}
In this section, we compare our group cohomology construction of FSPT drives to other approaches used in the literature.

An alternative approach to the classification of FSPT phases given in Refs.~\cite{Else:2016ja,Potter:2016tba} is based on studying the bulk eigenstate order of closed systems. In this approach, one maps the problem onto the classification of eigenstate order in a static system with symmetry $G\times\zbb$, which may be interpreted as $G\times \zbb_n$ for large enough $n$ that $H^{d+1}[G\times\zbb_n,{\rm U(1)}]$ is equal to $H^{d+1}[G,{\rm U(1)}]\times H^{d}[G,{\rm U(1)}]$. One then regards the bulk Floquet unitary operator $U$ as a symmetry operator corresponding to discrete time translation. While $U$ does not have the same form as the on-site symmetries of $G$, it was argued in Ref.~\cite{Else:2016ja} that, due to the finite speed of information flow, $U$ can nevertheless be treated on equal footing with the other discrete symmetries. The effective static symmetry group is then enhanced to $G\times \zbb$.

Ref.~\cite{Else:2016ja} proposed a set of nontrivial two-part drives, which fall into this paradigm, whose unitaries take the form $U=e^{-iH_{\rm SPT}}V(T)$. These consist of a drive with a trivial Hamiltonian, whose unitary corresponds to the action of a translation symmetry $V(T)$, followed by a drive with an SPT Hamiltonian $H_{\rm SPT}$ with group symmetry $G\times \zbb$. (Note that this is also the form of the 1d Floquet drives proposed in Ref.~\cite{vonKeyserlingk:2016bq}). It was then argued that when the operator $U$ is continuously deformed to $V(T)$, the group cohomology classification cannot change and is therefore given by $H^{d+1}[G\times\zbb,{\rm U(1)}]$. 

In our approach, the bulk eigenstate order of a Floquet system is independent of its dynamical order, which seems to be at odds with the approach outlined above. This motivates us to consider preceding the unitary drives above with one of the nontrivial loops introduced in Sec.~\ref{sec:cohomology_anyd}. For a closed system, the new Floquet unitary in the bulk is the same as the old unitary $U$. However, from the perspective adopted in this paper, this drive will be different from the drive without the nontrivial loop prepended. This shows by explicit construction that the choice of $U$ (as expressed in terms of local operators) is not unique. For example, consider a trivial $H_{\rm SPT}$ and prepend the unitary evolution $U=e^{-iH_{\rm SPT}}V(T)$ with a nontrivial loop; in the bulk, the unitary may still be chosen to be $U=e^{-iH_{\rm SPT}}V(T)$, which would lead to classification as a trivial phase. On the other hand, from the perspective of this paper, this is a nontrivial drive. 

For static systems there is always a unique choice of symmetry operator, and this choice does not depend on the boundary conditions involved. In contrast, for the driven systems that we study here, there is not a unique choice for the local decomposition of the time translation operator $U$ in a closed system. This choice could be fixed by considering its action on an open system, but such a choice does not seem to be continuously deformable to $V(T)$.

We note, however, the commonalities in predictions as well as physical interpretation between these approaches. Notably, the group cohomological part of the classification of FSPTs agrees with our approach based on loops, with the $H^d[G,{\rm U(1)}]$ part of the product $H^{d+1}[G\times\zbb,{\rm U(1)}]=H^{d+1}[G,{\rm U(1)}]\times H^{d}[G,{\rm U(1)}]$ (obtained using a K\"{u}nneth formula \cite{Cheng:2015vka}) seeming to account for the classification of loops. 

Another common aspect is the interpretation in terms of pumping for the two-dimensional case \cite{Else:2016ja,Potter:2016vq}. In Ref.~\cite{Else:2016ja}, it was argued that by gauging the full symmetry group, $\tilde{G}=G\times\zbb$, the effect of $U$ was the same as introducing a closed `symmetry-twist line', which is analogous to a 1d SPT phase. This was also a feature of the 1d classification schemes \cite{vonKeyserlingk:2016bq,Else:2016ja,Roy:2016ka,Potter:2016tba}. An explicit construction of a family of such drives in two dimensions was recently provided in Ref.~\cite{Potter:2016vq} for Abelian symmetry groups. In our current work, we find an explicit notion of pumping related to generating nontrivial SPT states at the edge of the system, extended to all dimensions, but we do not necessarily regard them as emerging from the bulk as in Refs.~\cite{Else:2016ja,Potter:2016tba,Potter:2016vq}.

While comparing approaches, we note that since we consider a homotopic classification of loops, the topology of a drive can be disentangled from the MBL restrictions of the endpoint unitaries. Thus, we are not restricted to cases where the eigenstates of the drive correspond to some fully MBL Hamiltonian.

\end{document}